\documentclass[a4paper,11pt]{article}

\usepackage{subcaption} % To allow subfigures
\usepackage{jcappub} % For JCAP, contains already hyperref, color, graphicx and many others
\usepackage{aas_macros} % To read NASA/ADS bibtex
\usepackage{physics} % A lot of useful macros to write equations in physics
\usepackage{cleveref} % Clever references, eg \cref{eq:eq_1} -> eq. (1), \cref{sec:sec_1} -> sec. 1.
\usepackage[normalem]{ulem}
\usepackage{xspace}

\newcommand{\Refs}[1]{refs.~{\cite{#1}}}
\newcommand{\Mp}{M_{\scriptscriptstyle{\mathrm{Pl}}}}

\newcommand{\efolds}{$e$-folds\xspace}
\newcommand{\Pfpt}[1]{P_{\scriptscriptstyle{\mathrm{FPT}} , #1}}
\newcommand{\PfptV}[1]{P^{\mathrm{V}}_{\scriptscriptstyle{\mathrm{FPT}} , #1}}

\subheader{}
\title{Compaction function in stochastic inflation: a \texttt{FOREST} of type I and II primordial black holes}

\author[a]{Chiara Animali,}
\author[b]{Pierre Auclair,}
\author[a]{Baptiste Blachier,}
\author[a]{Eemeli Tomberg,}
\author[c]{Vincent Vennin}

\affiliation[a]{Cosmology, Universe and Relativity at Louvain (CURL), Institute of Mathematics and
	Physics, University of Louvain, 2 Chemin du Cyclotron, 1348 Louvain-la-Neuve, Belgium}
\affiliation[b]{Sorbonne Université, CNRS, UMR 7095, Institut d'Astrophysique de Paris, 98 bis bd Arago, 75014 Paris, France}

\affiliation[c]{Laboratoire de Physique de l'\'Ecole Normale Sup\'erieure, ENS, CNRS, Universit\'e PSL, Sorbonne Universit\'e, Universit\'e Paris Cit\'e, F-75005 Paris, France}

\emailAdd{chiara.animali@uclouvain.be}
\emailAdd{auclair@iap.fr}
\emailAdd{baptiste.blachier@uclouvain.be}
\emailAdd{eemeli.tomberg@uclouvain.be}
\emailAdd{vincent.vennin@phys.ens.fr}

\date{today}

\graphicspath{{./figs/}}

\begin{document}
\sloppy

\abstract{We show how to compute the compaction function within stochastic inflation, by solving the random field dynamics on stochastic binary trees. In this framework, the compaction function is directly related to the ratio of the volumes emerging from the sibling and child branches of a given node. This construction also determines whether or not the areal radius of a perturbation increases monotonically with the radial coordinate, thereby distinguishing between type-I and type-II fluctuations. As an application, we investigate primordial black hole (PBH) formation in a single-field toy model with a constant potential slope, using stochastic-tree realizations generated with the public code \texttt{FOREST}. In the classical regime, where quantum diffusion is subdominant, the PBH mass function is narrowly distributed and type-II fluctuations are strongly suppressed relative to type I. By contrast, in the quantum and near-critical (i.e.\ close to eternal inflation) regimes, the PBH mass distribution spans several orders of magnitude, the overall PBH abundance is enhanced, and type-II fluctuations outnumber type I. In that case cloud-in-cloud effects are also important, highlighting the need for a better understanding of the evolution and collapse of type-II fluctuations in order to obtain robust PBH predictions when stochastic effects are significant.}

\maketitle

\section{Introduction}
\label{sec:intro}
While current cosmological observations provide strong bounds on the inflationary dynamics --- the inflaton should be semiclassical on the observed length scales, featuring Gaussian and nearly scale-invariant fluctuations~\cite{SDSS:2005xqv, Planck:2018nkj, Ivanov:2019pdj} --- periods of quantum diffusion may have affected the nature of fluctuations on small scales. In a regime where quantum fluctuations dominate over the classical dynamics, large and rare perturbations may have been produced, sourcing the formation of extreme objects such as primordial black holes (PBHs)~\cite{Hawking:1971ei, Carr:1974nx, Carr:1975qj}. A prototypical example model consists of a short, (nearly) flat domain in the inflaton potential, located towards the end of the semiclassical regime \cite{Pattison:2017mbe, Ezquiaga:2019ftu}.

In order to model the strongly inhomogeneous spacetime arising from such periods of quantum diffusion, non-perturbative techniques are required. A suitable non-perturbative framework is stochastic inflation~\cite{Starobinsky:1986fx}. It relies on the separate-universe picture~\cite{Salopek:1990jq, Sasaki:1995aw, Wands:2000dp, Lyth:2003im, Rigopoulos:2003ak, Lyth:2005fi, Artigas:2021zdk, Jackson:2023obv}, in which the inflating universe is viewed as an ensemble of independent Hubble-sized patches that are each locally homogeneous and isotropic. The evolution in each patch is then described by a Langevin equation for the coarse-grained inflaton field. The stochasticity arises from short-wavelength quantum fluctuations that cross the Hubble radius during inflation and subsequently affect the local background evolution. The cumulated contribution of such sub-Hubble fluctuations takes the form of a Gaussian white noise $\xi$ acting on the dynamics of the coarse-grained, super-Hubble inflaton field $\phi$. In each Hubble-sized patch, the dynamics is thus governed by a Langevin equation
\begin{equation}
	\label{eq:Langevin}
	\dv{\phi}{N} = -\frac{V'(\phi)}{3H^2} + \frac{H}{2\pi}\xi(N)
	\qq{with} H
    ^2=\frac{V(\phi)}{3\Mp^2} \, ,
\end{equation}
valid for single-field slow-roll models~\cite{Vennin:2015hra, Lesgourgues:1996jc, Grain:2017dqa}. 
The statistical properties of the curvature perturbation $\zeta$ are then inferred using a stochastic version of the $\delta N$ formalism~\cite{Starobinsky:1982ee, Starobinsky:1985ibc, Sasaki:1995aw, Sasaki:1998ug, Lyth:2004gb, Lyth:2005fi, Finelli:2008zg, Finelli:2010sh}. The latter promotes each patch's local amount of expansion $N(t,\vec{x})=\ln[a(t,\vec{x})]$, where $a(t,\vec{x})$ stands as the local value of the scale factor, to a stochastic quantity~\cite{Enqvist:2008kt, Fujita:2013cna, Vennin:2015hra, Pattison:2017mbe}. The curvature perturbation at the end of inflation is thus inferred as
\begin{equation}
    \zeta(\vec{x})= \mathcal{N}(\vec{x})-\langle \mathcal{N}\rangle \equiv \delta N\,,
\end{equation}
where $\mathcal{N}(\vec{x})$ is the number of \efolds until the end of inflation realized along the stochastic trajectory ending at $\vec{x}$, and $\langle\mathcal{N}\rangle$ is its average. In particular, it has been shown that exponential tails are a generic outcome for the distribution of large curvature fluctuations~\cite{Pattison:2017mbe,Ezquiaga:2019ftu,Panagopoulos:2019ail,Figueroa:2020jkf,Figueroa:2021zah,Pattison:2021oen,Tomberg:2021xxv, Rigopoulos:2021nhv, Achucarro:2021pdh,Ezquiaga:2022qpw, Animali:2022otk, Cai:2022erk, Gow:2022jfb, Tomberg:2023kli, Briaud:2023eae, Vennin:2024yzl, Inui:2024sce, Sharma:2024fbr}.

Accessing the real-space profile of the curvature fluctuation or other quantities derived from it is challenging within the stochastic formalism. To go beyond dedicated (semi)-analytical approximation schemes~\cite{Ando:2020fjm, Tada:2021zzj, Animali:2024jiz}, ref.~\cite{Animali:2025pyf} conceptualized the inflationary expansion as a branching process and implemented the stochastic-$\delta N$ formalism on stochastic trees~\cite{Linde:1993xx,Jain:2019gsq}. The key idea is that the inflationary spacetime can be framed as a branching tree, whose nodes correspond to Hubble patches. As the comoving region encapsulated in a given node expands, its physical volume grows until it doubles, giving rise to two new nodes. 

These nodes, each representing a descendant Hubble patch, subsequently follow two independent Langevin realizations. This branching is the elementary building block of the stochastic tree, and it repeats recursively, until the inflaton field value reaches the end of inflation (or the end of the region with strong quantum diffusion) in each of the patches. The set of these terminal nodes, the ``leaves'' of the tree, defines the final hypersurface on which cosmological fields, such as the curvature perturbation, are measured. Accordingly, the statistical properties of these fields are encoded in the tree structure and can be extracted from it.

By simulating stochastic trees via the numerical code \texttt{FOREST}~\cite{auclair_2025_15235932}, in ref.~\cite{Animali:2025pyf} it was shown how to obtain various statistical quantities such as the (end-of-inflation) volume and the duration of inflation (first-passage-time) distributions, as well as real-space maps of the inhomogeneous end-of-inflation hypersurface generated by quantum diffusion. In particular, stochastic trees turned out to be well-suited for harvesting PBHs, inferring both their abundance and mass, while automatically taking into account the cloud-in-cloud phenomenon~\cite{Jedamzik:1994nr}, i.e.\ the fact that a hierarchy of overdense regions can lead to PBH formation inside larger PBHs. 

The criterion for PBH formation used in ref.~\cite{Animali:2025pyf} relied on the reconstruction of the \emph{coarse-shelled curvature perturbation}, namely, the curvature perturbation averaged between two concentric spheres~\cite{Tada:2021zzj}. Although the quantity that characterizes the threshold for PBH formation has been the subject of much debate~\cite{Shibata:1999zs, Niemeyer:1997mt, Musco:2004ak, Nakama:2013ica, Harada:2013epa}, the standard practice relies on a quantity called the \emph{compaction function}~\cite{Shibata:1999zs}. The compaction function's peak value at super-Hubble scales predicts whether positive amplitude fluctuations will collapse into a black hole upon horizon re-entry, an observation that has been confirmed in several numerical works in various scenarios~\cite{Harada:2015yda, Musco:2018rwt, Escriva:2019phb, Musco:2020jjb, Escriva:2020tak}. As a consequence, realistic estimates of the PBH abundance should account for the radial profile of curvature fluctuations and the corresponding compaction function. The coarse-shelled curvature perturbation was designed to serve as a proxy for the compaction function, but did not provide access to its radial profile. Moreover, when the PBH formation criterion was imposed on it, it turned out to be highly sensitive to volume imbalances within the tree, leading to suspicious upward propagation of PBH-forming regions towards larger scales. As a result, the predicted PBH population would sometimes become dominated by a few very massive black holes, or even by a single one in the most extreme cases. The purpose of the present work is thus to refine that analysis by fully reconstructing the curvature and compaction-function profiles within the stochastic-tree framework.

Compaction-function profiles were previously studied in stochastic constant-roll inflation in \Refs{Raatikainen:2023bzk, Raatikainen:2025gpd}. Here, we provide an alternative implementation in a different setup which proves to be very efficient. 
Indeed, in a stochastic tree, the procedure boils down to computing ratios of volumes associated with nodes. Because the stochastic-tree framework is naturally formulated in terms of volumes, these are among the most readily accessible quantities in simulations. In addition, unlike the coarse-shelled proxy, it enables us to disambiguate between type-I and type-II perturbations~\cite{Kopp:2010sh, Carr:2014pga, Escriva:2023uko, Harada:2024jxl, Uehara:2024yyp, Escriva:2025eqc, Escriva:2025rja, Inui:2024fgk, Uehara:2025idq, Escriva:2025ftp}, which could give rise to different types of PBHs upon collapse, and to predict how the competition between the classical dynamics and the quantum diffusion affects their abundance.

The article is organized as follows. In \cref{sec:compact}, we briefly review the compaction function formalism and explain how to implement it within stochastic trees. This requires choosing an appropriate discretization scheme and specifying a mapping that locally embeds the stochastic-tree topology into real space. In \cref{sec:pbh}, we show how the reconstructed compaction function can be used to identify primordial black holes sourced by both type-I and type-II curvature perturbations. We discuss the assumptions and the criteria used to delineate the type-I/II regions and to assign masses to the corresponding PBHs. In \cref{sec:application}, we apply these methods to the ``tilted quantum-well'' toy model of inflation, which exhibits both quantum diffusion and a well-defined classical limit. Using a population of $10^9$ to $10^{10}$ stochastic trees, we reconstruct the mass fraction and mass distribution of PBHs for type-I and type-II perturbations. In \cref{sec:discussion}, we address several technical aspects of the discretization scheme and compare our results with earlier analytical approximations and previously used frameworks. Finally, in \cref{sec:conclusion}, we summarize our results and highlight prospects for future work.

\section{Compaction function in stochastic inflation}
\label{sec:compact}

\subsection{Compaction function}

We consider the scenario where PBHs originate from the collapse of curvature fluctuations $\zeta$ produced during the inflationary epoch and stretched to super-Hubble scales, as they re-enter the Hubble radius afterwards. In the gauge where fixed $t$ slices have uniform energy density, and where fixed $\vec{x}$ worldlines are comoving (this coincides with the synchronous gauge at super-Hubble scales~\cite{Salopek:1990jq}), at super-Hubble scales the spacetime metric takes the form 
\begin{equation}
    \dd s^2 = -\dd t^2 + a^2(t)e^{2\zeta(r)} \qty(\dd r^2 + r^2 \dd \Omega^2) \, ,
\label{eq:super_hubble_metric}
\end{equation}
where $\dd \Omega^2 = \dd \theta^2 + \sin^2 \theta \, \dd \phi^2$ and we have assumed that, near its peak, the curvature perturbation follows a spherically symmetric profile and only depends on the comoving radial coordinate $r$.
This provides a good approximation for high peaks leading to PBH formation, at least when they stem from Gaussian fluctuations~\cite{Bardeen:1986} (moreover, the resulting gravitational collapse also tends to be spherical~\cite{Escriva:2024aeo, Escriva:2024lmm}).

The curvature perturbation $\zeta$ in \cref{eq:super_hubble_metric} is defined with respect to a background encompassing the whole observable universe. However, gravitational collapse is a local process, and PBH formation should therefore be insensitive to averages taken over regions much larger than the Hubble radius at the time of the collapse. This indicates that $\zeta$ is not the most suitable quantity for assessing whether a given patch will collapse. A more appropriate diagnostic is instead provided by the compaction function, defined as~\cite{Shibata:1999zs}
\begin{equation}
    \mathcal{C}(r,t) \equiv \frac{2 \delta M(r,t)}{R(r,t)}\, ,
\end{equation}
where
\begin{equation}
    R(r,t) \equiv a(t) e^{\zeta(r)} r \ 
\label{eq:aerialR}
\end{equation}
is the areal radius, so that a sphere at fixed $r$ and $t$ has area $4\pi R^2(r,t)$ and $\delta M(t)$ is the mass excess inside that sphere, defined as the difference between the Misner-Sharp mass and the background mass within the same areal radius calculated with respect to a local Friedmann-Lema\^{i}tre-Robertson-Walker (FLRW) universe. 

At leading order in the gradient expansion, and at a fixed time, one can express the compaction function solely in terms of the curvature perturbation as~\cite{Harada:2015yda}
\begin{equation}
	\mathcal{C}(r)=z(w)\left\{ 1-\left[1+r\zeta'(r)\right]^2\right\} \, , \qquad
	z(w) \equiv \frac{3(1+w)}{5+3w} \, ,
\label{eq:compaction-func}
\end{equation}
with $w$ the equation-of-state parameter. As is clear in the above equation, the compaction function is a fully non-linear quantity. Nevertheless, as discussed for example in refs.~\cite{Germani:2023ojx,Escriva:2025rja}, its properties can be described equivalently, and more conveniently, in terms of the associated linear compaction function~\cite{Germani:2019zez,Germani:2023ojx},
\begin{equation}
    \mathcal{C}_\mathrm{l}(r) \equiv -2 z(w) r \zeta'(r) \quad \implies \quad
    \mathcal{C}(r) = \mathcal{C}_\mathrm{l}(r) - \frac{\mathcal{C}_{\mathrm{l}}^{2}(r)}{4 z(w)} \, .
\label{eq:lin-compact}
\end{equation}
On super-Hubble scales, $\mathcal{C}_\mathrm{l}(r)$ has a simple physical interpretation as it coincides with the comoving density contrast smoothed over a sphere of comoving radius $r$, as follows from the linear Poisson equation in a flat FLRW universe.

Differentiating \cref{eq:aerialR} at fixed time gives
\begin{equation}
    \frac{\dd \ln R}{\dd \ln r} = 1 + r\zeta'(r) = 1 - \frac{\mathcal{C}_\mathrm{l}(r)}{2z(w)} \, .
\label{eq:dlnR}
\end{equation}
Thus, knowledge of the areal-radius profile $R(r)$ is sufficient to determine the full compaction-function profile $\mathcal{C}(r)$. 

In the implementation of stochastic inflation on stochastic trees developed in ref.~\cite{Animali:2025pyf}, smoothing is implemented at the level of the curvature perturbation, leading to the definition of a coarse-shelled curvature perturbation~\cite{Tada:2021zzj}. A correspondence between this quantity and the linear compaction function is then established by matching the corresponding window functions. Such a procedure, however, is only designed as a proxy, and does not fully reproduce the compaction function, which is arguably the key quantity for PBH formation. In what follows, we show that the full compaction function in \cref{eq:compaction-func} can in fact be computed within the stochastic-tree framework, using the relation between the areal radius $R(r)$ and $\mathcal{C}_\mathrm{l}(r)$, see \cref{eq:dlnR}. A quantitative comparison between the two methods is presented in \cref{subsec:comp:coarse:shelled}.

\subsection{Compaction function from stochastic trees}
\label{sec:compact:trees}

We begin by briefly reviewing the ideas underlying the stochastic-tree framework introduced in ref.~\cite{Animali:2025pyf}. Within a stochastic tree, the inflaton field $\phi$ is averaged over a region of physical size $R_\sigma=(\sigma H)^{-1}$, where $\sigma\ll 1$ is a fixed parameter that sets the ratio between the coarse-graining radius $R_\sigma$ and the Hubble radius $H^{-1}$. We assume $H$ to be approximately constant during inflation, so that $R_\sigma$ defines a fixed coarse-graining scale. As inflation proceeds and space expands, a coarse-grained coordinate patch grows in physical volume as $a^3(t) \propto e^{3N}$, so that the volume doubles after $\Delta N=\ln(2)/3$ \efolds. A parent patch thus gives rise to two child patches with a shared past but no future causal contact, thereby realizing the separate-universe picture underlying stochastic inflation on trees. In practice, this implies that the field values in these two patches are evolved forward from a common value $\phi_i$, that of the parent patch at the moment of the splitting, using two distinct realizations of the Langevin equation~\eqref{eq:Langevin}. The child patches still have physical size given by $R_\sigma$, but their comoving volumes are twice smaller, hence
\begin{equation}
	\Delta \ln r = -\ln(2)/3  =- \Delta N
\label{eq:step}
\end{equation}
between the splittings.

\begin{figure}
    \centering
    \includegraphics{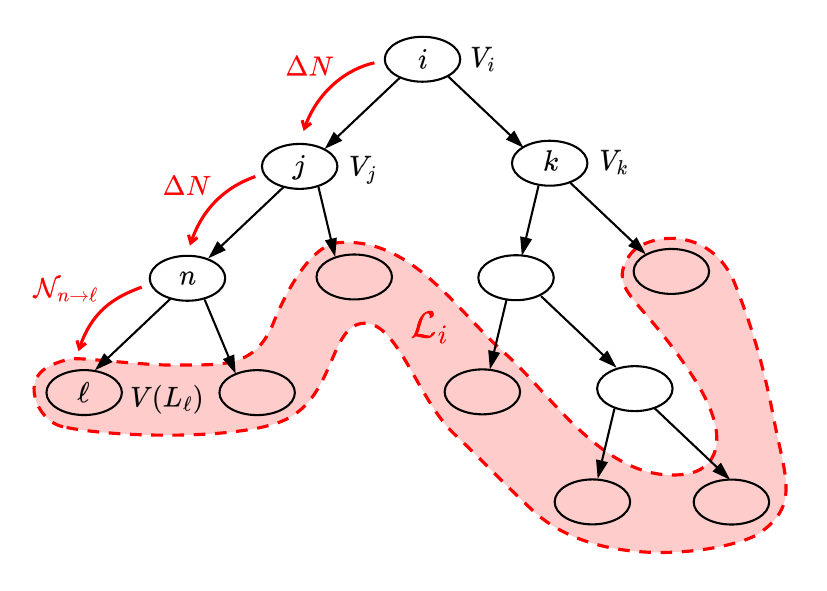}
    \caption{An example stochastic tree, depicting branching nodes ($i,j,k,\dots$), the \efolds\ between two consecutive bulk nodes ($\Delta N$) and from the last bulk node to a leaf ($\mathcal{N}_{n\to\ell}$), the full set of leaves descending from node $i$ ($\mathcal{L}_i$), the volume emerging from branching nodes ($V_i, V_j, \dots$) and the volume of individual leaves $V(L_\ell)$.}
    \label{fig:example_tree}
\end{figure}

By iterating such an ``elementary vertex'' one obtains a binary-tree structure, which we call the stochastic tree (see \cref{fig:example_tree} for an example). As the end of inflation is reached in each independent patch, the tree stops growing, and the final nodes where inflation stops correspond to the ``leaves'' of the tree. In single-field slow-roll models, this occurs by slow-roll violation, i.e.\ when $\phi$ reaches a value $\phi_\mathrm{end}$ where the potential function becomes too steep for inflation to proceed. In units of the reference volume $V_\sigma=4\pi R_\sigma^3/3$,
a leaf's (denoted by $L_\ell$) physical volume is given by
\begin{equation}
	\label{eq:volume:leaf}
	V\left(L_\ell\right) = e^{3 \left(\mathcal{N}_{i\to \ell}-\Delta N\right)} \in [1/2, 1[ \, ,
\end{equation}
where $\mathcal{N}_{i\to \ell} < \Delta N$ denotes the number of \efolds realized along the final ``branch'' connecting the parent patch $i$ to the leaf $\ell$. In practice, we adopt a split-then-grow prescription for the branching: as soon as a patch reaches the reference physical volume $V_\sigma$, it is replaced by two child patches. Because the parent patch is still spatially flat at the splitting time, the children are born with equal comoving volumes, each equal to one half of the parent’s, and hence with physical volume $V_\sigma/2$. They then grow until either they split again or inflation ends. This is reflected in the range of possible leaf volumes in \cref{eq:volume:leaf}. Most leaves do not reach the size of a Hubble patch, in contrast to the node-patches in the ``bulk'' of the tree.

Let us now consider the volume emerging from each node of the tree. Starting from the leaves, the volume propagates upwards, so that, when a node $i$ gives rise to two child nodes $j$ and $k$, we have
\begin{equation}
	\label{eq:iter:V}
	V_i=V_j+V_k\,.
\end{equation}
If $\mathcal{L}_i$ denotes the set of leaves descending from the root node $i$, the emerging volume $V_i$ is thus the sum of the volumes $V(L_\ell)$ of the leaves in $\mathcal{L}_i$,
\begin{equation}\label{eq:Vi}
	V_i = \sum_{\ell \in \mathcal{L}_i} V(L_\ell) \, .
\end{equation}

Having explained how the volume associated with branching nodes and leaves can be computed from a stochastic tree, we now turn to the calculation of the compaction function \eqref{eq:compaction-func} within this framework. To this end, we first relate the volumes of the nodes to the areal radius, and subsequently to the compaction function. Let us consider the volume element of a spherical shell at radius $r$, with thickness $\dd r$, in the metric \eqref{eq:super_hubble_metric},
\begin{equation}
    \dd V = 4\pi a^3e^{3\zeta(r)}r^2 \dd r
    = 4\pi R^3(r) \dd \ln r \, ,
\label{eq:dV}
\end{equation}
where the angular coordinates have been integrated out. This gives the areal radius $R(r)$ as
\begin{equation}
    R^3(r) = \frac{1}{4\pi}\frac{\dd V}{\dd \ln r} \, .
\label{eq:R3}
\end{equation}
Using \cref{eq:dlnR}, we obtain
\begin{equation}
    \mathcal{C}_\mathrm{l} (r) = 2z(w)\qty[1 - \frac{1}{3}\frac{\dd}{\dd \ln r}\ln(\frac{\dd V}{\dd \ln r})] \, .
\label{eq:C_in_V}
\end{equation}

\begin{figure}
    \centering
    \includegraphics{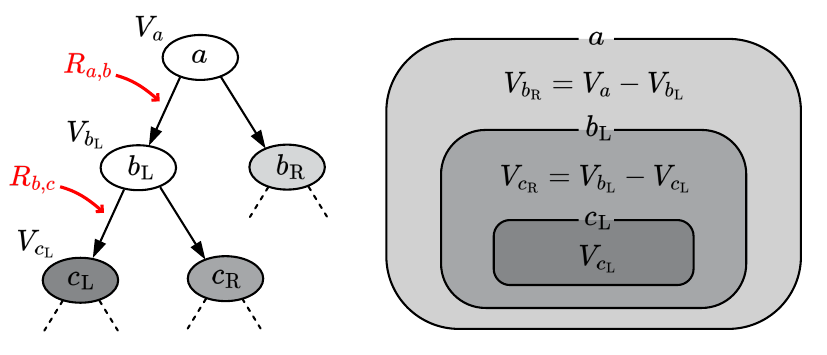}
    \caption{The three generations of nodes used in \cref{eq:R_discrete,eq:Cl_discrete}, together with the fiducial mapping of their volumes into real space.}
    \label{fig:nodes_in_derivatives}
\end{figure}

By construction, the stochastic tree only gives access to a discrete set of smoothing scales. 
The derivatives appearing in eqs.~\eqref{eq:R3} and~\eqref{eq:C_in_V} must therefore be discretized accordingly. The second derivative in \cref{eq:C_in_V} requires information from three consecutive nodes along a branch: the node around which we perform the computation, its parent, and one of its child nodes. Above, we chose to express derivatives with respect to $\ln r$; this is because, between two consecutive nodes, $\Delta \ln r = -\ln(2)/3$ is constant, see \cref{eq:step}.\footnote{The segments preceding terminal leaves do not necessarily have $\Delta N=\ln(2)/3$, but are always such that $\Delta\ln r=-\ln(2)/3$ (comoving volume is split in two at branching), which is the only thing that matters when discretising \cref{eq:R3}. 
\label{footnote:DeltaNversusDeltalnr}
}
This allows us to discretize \cref{eq:R3} as
\begin{equation}
    R^3_{a,b} = \frac{3}{4\pi\ln(2)}\qty(V_a - V_{b_\mathrm{L}}) \, , \qquad
    R^3_{b,c} = \frac{3}{4\pi\ln(2)}\qty(V_{b_\mathrm{L}} - V_{c_\mathrm{L}}) \, ,
    \label{eq:R_discrete}
\end{equation}
and \cref{eq:C_in_V} as
\begin{equation}
    \mathcal{C}_{\mathrm{l}, b_\mathrm{L}} = 2z(w)\qty[1 - \frac{\ln(R^3_{a,b}) - \ln (R^3_{b,c})}{\ln(2)}]
    = 2z(w)\left[ 1 - \log_2 \left( \frac{V_a - V_{b_\mathrm{L}}}{V_{b_\mathrm{L}} - V_{c_\mathrm{L}}}\right) \right] \, .
\label{eq:Cl_discrete}
\end{equation}
Here $b_\mathrm{L}$ is the node under inspection, $a$ is its parent node, and $c_\mathrm{L}$ is the child node. The areal radii are computed between consecutive nodes, that is, they correspond to radial coordinates $r$ that are intermediate between those of the nodes under consideration. We depict these nodes in the left panel of \cref{fig:nodes_in_derivatives}, together with the sibling node $b_{\mathrm{R}}$ and the other child node $c_\mathrm{R}$. In practice, using the additivity of the volumes in~\cref{eq:iter:V}, the discretized linear compaction function can be rewritten solely in terms of the volumes of the off-branch nodes $b_\mathrm{R}$ and $c_\mathrm{R}$ according to
\begin{equation}
    \mathcal{C}_{\mathrm{l}, b_\mathrm{L}} = 2z(w)\left[ 1 - \log_2 \left( \frac{V_{b_\mathrm{R}}}{V_{c_\mathrm{R}}}\right) \right] \, .
\label{eq:Cl_offbranch}
\end{equation}
This expression highlights the efficiency of the proposed implementation of the compaction function. Up to constant prefactors, it is formulated entirely in terms of the node volumes, namely the physical extensive quantities that are most directly accessible in stochastic trees.

Finally, although there are other ways to discretize \cref{eq:R3,eq:C_in_V}, the scheme adopted above exhibits a number of desirable properties that ensure consistency with the continuum limit. In particular, if the sibling branches have the same volume in each generation, then $V_{b_{\mathrm{R}}}=V_{b_{\mathrm{L}}}=2V_{c_{\mathrm{R}}}$ and \cref{eq:Cl_offbranch} gives $\mathcal{C}_{\mathrm{l},b_{\mathrm{L}}}=0$. This is consistent with the fact that such a ``maximally balanced'' (sub)tree carries no perturbation. We will further discuss balancing and other properties of the discretization scheme in \cref{subsec:discretization:scheme} (see also footnote~\ref{footnote:typeII:criterion:discretized} below).

\section{Harvesting primordial black holes}
\label{sec:pbh}

\subsection{Type-I and type-II regions}
\label{sec:pbh:perturb_I_II}

\begin{figure}
	\begin{subfigure}{.325\textwidth}
		\includegraphics[width=\textwidth]{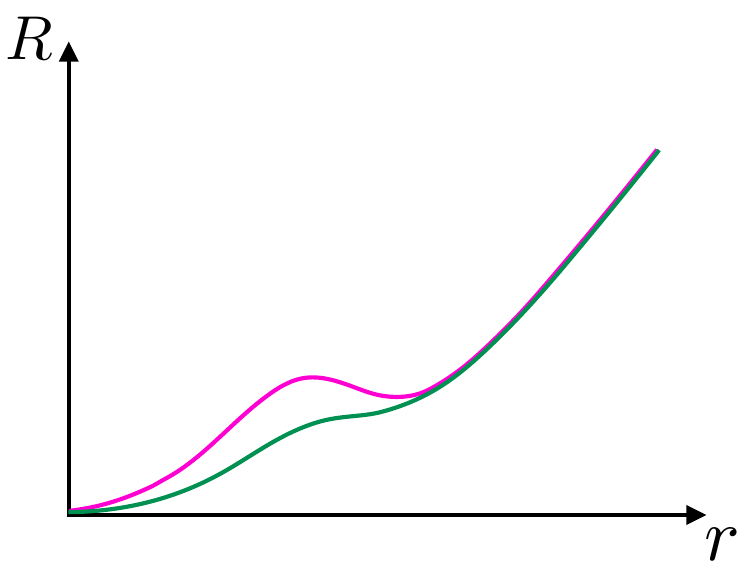}
		\caption{Areal radius}
	\end{subfigure}
	\begin{subfigure}{.325\textwidth}
		\includegraphics[width=\textwidth]{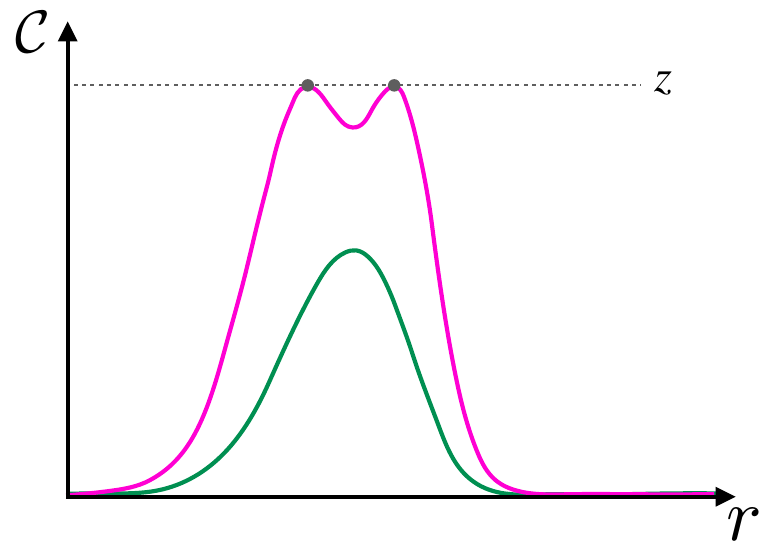}
		\caption{Compaction}
	\end{subfigure}
	\begin{subfigure}{.325\textwidth}
		\includegraphics[width=\textwidth]{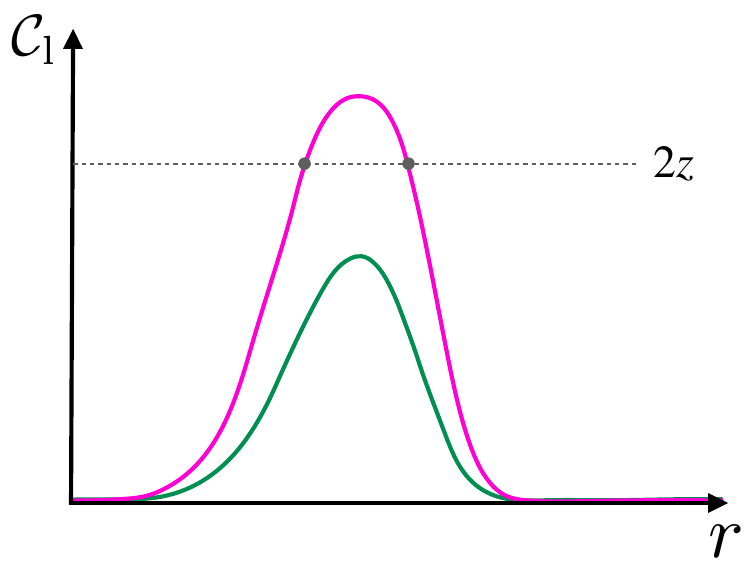}
		\caption{Linear compaction}
	\end{subfigure}
	\caption{Schematic illustration of the behaviour of the areal radius (a), compaction function (b) and linear compaction function (c) as functions of the radial coordinate $r$ for type-I (green) and type-II (magenta) fluctuations.
	}
	\label{fig:compaction:sketch}
\end{figure}

As first pointed out in ref.~\cite{Shibata:1999zs}, the maximum of the compaction function is the relevant quantity for characterizing the formation threshold of primordial black holes. Nevertheless, the local shape of the curvature perturbation may also affect the collapse process and the type of PBH being formed. We now discuss how this proceeds, following the situation sketched in \cref{fig:compaction:sketch}.

Let us denote by $r_\mathrm{m}$ the radius at which the linear compaction function $\mathcal{C}_\mathrm{l}$ reaches its maximum value. Upon differentiating \cref{eq:lin-compact} with respect to $r$, one obtains 
\begin{equation}
    \mathcal{C}'(r) = \mathcal{C}_{\mathrm{l}}'(r)  \frac{\dd \ln R}{\dd \ln r} \, ,
\label{eq:dcompact}
\end{equation}
where we have also used \cref{eq:dlnR}. In the simple case where fluctuations have an areal radius that increases monotonically with $r$ (green curves in \cref{fig:compaction:sketch}), i.e.\, $R'(r) > 0$, the compaction function inherits the same monotonicity as the linear compaction function. In particular, the maximum $\mathcal{C}_\mathrm{l}(r_\mathrm{m})$ then corresponds to a maximum $\mathcal{C}(r_\mathrm{m})$. Using \cref{eq:dlnR}, this holds whenever $\mathcal{C}_\mathrm{l} (r_\mathrm{m}) \leq 2z(w)$, which defines the so-called type-I region.

Some perturbations satisfy $\mathcal{C}_\mathrm{l} (r_\mathrm{m}) > 2z(w)$, corresponding to a type-II region. In this case, corresponding to the magenta curves in \cref{fig:compaction:sketch}, the areal radius is a non-monotonic function of $r$, with $R'(r) < 0$~\cite{Kopp:2010sh}, that is, the areal radius  locally \emph{decreases} as the coordinate radius increases.\footnote{This property is satisfied by our discretized scheme, \cref{eq:R_discrete,eq:Cl_discrete}, which guarantees that $\mathcal{C}_\mathrm{l}<2z\Leftrightarrow R_{a,b} < R_{b,c}$.\label{footnote:typeII:criterion:discretized}} From \cref{eq:dcompact}, it follows that the maximum of $\mathcal{C}_\mathrm{l}$ corresponds to a \emph{minimum} of $\mathcal{C}$. In practice, this minimum is always surrounded by two maxima~\cite{Uehara:2024yyp, Harada:2024trx}. Starting from large radii, where $\mathcal{C}=\mathcal{C}_\mathrm{l} = 0$ and as the radius decreases, the compaction function first increases and reaches a maximum value $\mathcal{C}=z(w)$, corresponding to $\mathcal{C}_\mathrm{l} = 2z(w)$. As the radius is decreased further and the profile enters the type-II region ($\mathcal{C}_\mathrm{l} > 2z(w)$), the compaction function decreases to a local minimum at $r_\mathrm{m}$. As the radius approaches zero, the behaviour of $\mathcal{C}$ reverses, increasing up to the second maximum and then decreasing again ($\mathcal{C}_\mathrm{l}$ returns to zero for $r \rightarrow 0$).

\subsection{PBH formation from type-I and type-II perturbations}
\label{sec:pbh:threshold_I_II}

The type-I/II classification refers to the \emph{geometry} of the super-Hubble fluctuation.\footnote{Another geometric classification, based on the sign of the spatial curvature at the fluctuation’s core, has been recently proposed in ref.~\cite{Germani:2025hcu}.} Separately, the PBHs that form can be classified according to their horizon structure into type A and B. This classification is \emph{dynamical}, as it depends on the collapse process, which is established through numerical simulations. Type-A PBHs correspond to the standard scenario in which overdensities collapse after crossing the Hubble radius and subsequently form an apparent horizon. In contrast, type-B PBHs form through a ``separate universe'' mechanism in which a sufficiently large curvature fluctuation generates a ``throat'' connecting a high-volume bubble of expanded space to its surrounding~\cite{Uehara:2024yyp, Shimada:2024eec}.

Type-I perturbations tend to form type-A primordial black holes, while type-II perturbations tend to form type-B primordial black holes. However, this correspondence is not perfect when pressure is non-zero ($w>0$) \cite{Harada:2024jxl, Uehara:2025idq}, as is the case during radiation domination considered below.
The stochastic approach only provides information about the curvature perturbation, and does not resolve the horizon structure. Therefore, throughout this paper, we adopt the PBH classification based on the perturbation type (I or II). This is, in fact, a general practice in the literature whenever qualitative estimates of PBH abundance are derived.

In standard scenarios, the contribution of type-II fluctuations to the total PBH abundance is often neglected, since such configurations are statistically rare.
However, as first pointed out in refs.~\cite{Gow:2022jfb, Escriva:2023uko}, if the statistics of the curvature fluctuations exhibit non-Gaussianities, the threshold of PBH formation may be ``pushed'' into the region of type-II fluctuations, thus enhancing the contribution of type-II PBHs. This was recently confirmed in refs.~\cite{Shimada:2024eec, Inui:2024fgk, Escriva:2025eqc} for large local-type non-Gaussianities and also, perhaps more surprisingly, in ref.~\cite{Fumagalli:2024kgg} for nearly Gaussian initial curvature perturbations that are enhanced across a broad range of scales.

The implementation of the compaction function in stochastic trees, detailed in \cref{sec:compact:trees}, allows us to assess this picture when quantum diffusion is taken into account. To this end, we need prescriptions for the collapse thresholds in the type-I/II regions. Recent work shows that in a radiation-dominated universe, the threshold on the compaction function $\mathcal{C}_\mathrm{c}$ lies in the range $ \mathcal{C}_\mathrm{c} \in \left[ 2/5, 2/3\right]$ in the type-I region, with the exact value depending on the shape of the compaction-function profile~\cite{Musco:2018rwt, Musco:2020jjb, Germani:2018jgr, Escriva:2020tak}. In the type-II region, collapse is better described in terms of a threshold $\mathcal{C}_{\mathrm{l},\mathrm{c}}$ for the linear component, again varying with the profile shape, with the lower bound $2z(w)$ (the edge of the type-II region) and no apparent upper bound~\cite{Escriva:2025eqc, Escriva:2025rja}.

Since stochastic trees provide direct access to the linear compaction function (see \cref{eq:Cl_discrete}), we define the PBH formation threshold in terms of $\mathcal{C}_\mathrm{l}$ throughout. As eqs.~\eqref{eq:lin-compact} and~\eqref{eq:compaction-func} show, there are two values of $\mathcal{C}_\mathrm{l}$ corresponding to a given $\mathcal{C}$, which we denote by
\begin{equation}
    \mathcal{C}_{\mathrm{l}}^{\pm}(\mathcal{C}) = 2z(w) \left[ 1 \pm \sqrt{1 - \frac{\mathcal{C}}{z(w)}}   \right] \, ,
\end{equation}
where $\mathcal{C}_{\mathrm{l}}^{-}$ lies in the type-I region and $\mathcal{C}_{\mathrm{l}}^{+}$ lies in the type-II region. In this work, we reconstruct the abundance of over-the-threshold type-I fluctuations, and of type-II fluctuations, calling them type-I and type-II ``PBHs'' respectively for simplicity. We remain however agnostic about the type of astrophysical objects they actually form (except when assigning a mass to these objects, which follows the procedure outlined in \cref{sec:PBH:mass}). Denoting $\mathcal{C}_\mathrm{l,c}=\mathcal{C}_\mathrm{l}^{-}(\mathcal{C}_\mathrm{c})$, where we recall that $\mathcal{C}_\mathrm{c}$ is the compaction-function threshold in the type-I region, we then distinguish three cases:
\begin{itemize}
    \item $\mathcal{C}_\mathrm{l} (r_\mathrm{m}) < \mathcal{C}_\mathrm{l,c}$: type-I fluctuations that do not collapse into PBHs;
    \item $\mathcal{C}_\mathrm{l,c} < \mathcal{C}_\mathrm{l}(r_\mathrm{m}) < 2z(w)$: type-I fluctuations that collapse upon Hubble horizon re-entry into ``type-I PBHs'';
    \item $\mathcal{C}_\mathrm{l} (r_\mathrm{m}) > 2 z(w)$: type-II fluctuations that collapse upon Hubble horizon re-entry into ``type-II PBHs''.
\end{itemize}

\subsection{Finding primordial black holes in stochastic trees}
\label{sec:BH:stoch:trees}

\begin{figure}
    \centering
    \includegraphics[width=.7\textwidth]{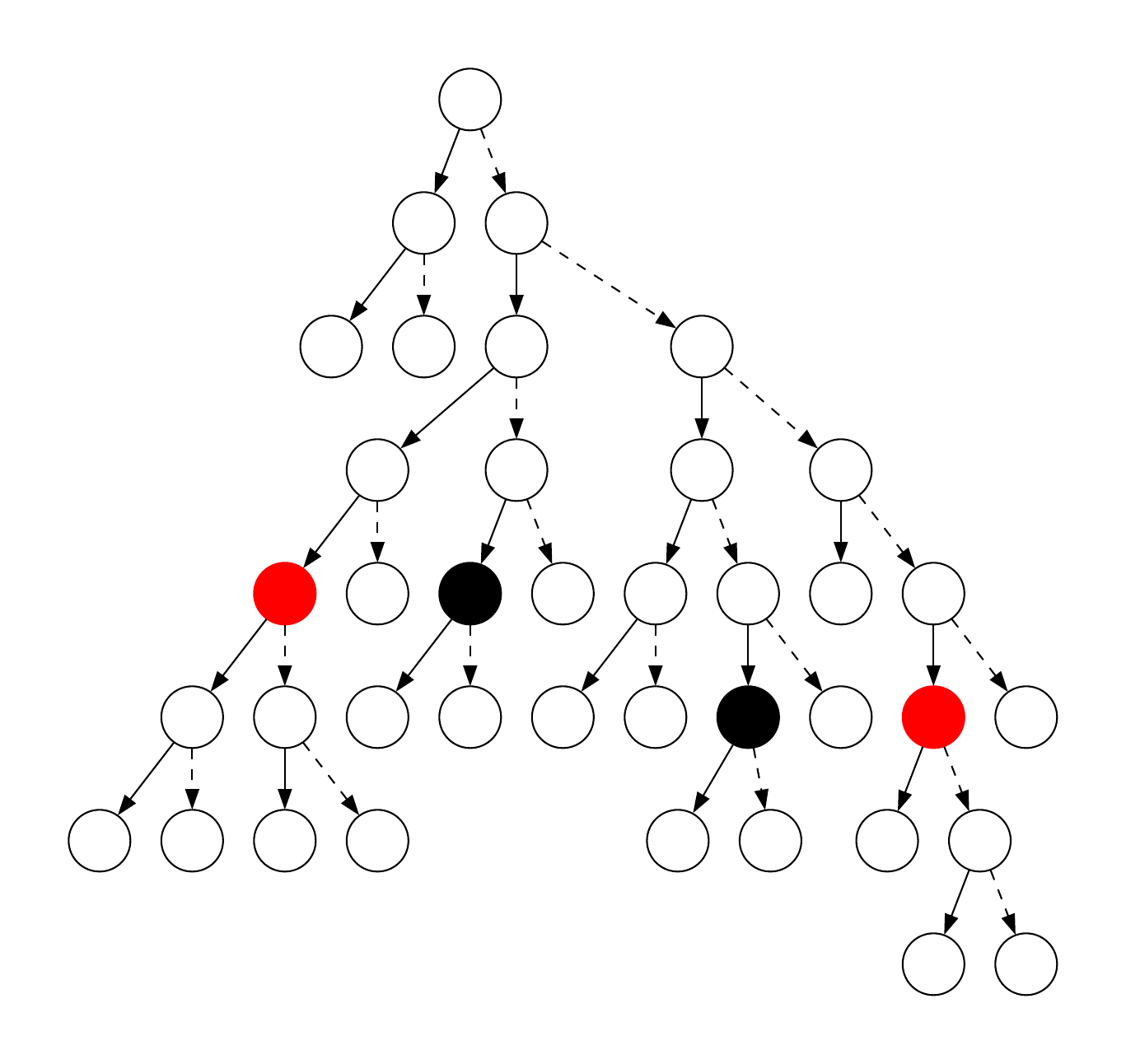}
    \caption{An example tree arising from the tilted-well potential of \cref{sec:application}. The black and red nodes correspond to type-I and II PBH-forming nodes, respectively. Solid arrows point to left-hand nodes and dashed arrows point to right-hand nodes.}
    \label{fig:numerical_example_tree}
\end{figure}

We now describe how primordial black holes are identified in practice within a stochastic tree, how their type is classified, and which assumptions we adopt in this procedure.

Any stochastic tree can be viewed as a collection of subtrees of the form displayed in \cref{fig:nodes_in_derivatives}. Adopting the ordering prescription explained in \cref{sec:compact:trees}, we compute the linear compaction function at each (left) bulk node using \cref{eq:Cl_discrete}. A node is then flagged as a PBH-forming one if the collapse criterion $\mathcal{C}_\mathrm{l} > \mathcal{C}_\mathrm{l,c}$ is satisfied.

From a numerical perspective, the code $\texttt{FOREST}$ is built on a recursive algorithm that iterates the elementary vertex until inflation ends at the tip of every branch. Quantities such as the compaction function are evaluated \emph{dynamically} at each node during the construction of the tree. Information including the volume of a node, the number of PBHs, their mass, etc, is propagated upwards and stored in the stack as the tree is climbed up to the root node. Hence, concomitantly with the creation of the tree, the ``off-branch'' volumes can be evaluated for each triplet of nodes. PBHs are then flagged as type I or type II depending on the value of $\mathcal{C}_\mathrm{l}$ at the PBH-forming node. This procedure can be carried out straightforwardly by adapting the formation criteria discussed in \cref{sec:pbh:threshold_I_II} to \cref{eq:Cl_offbranch}, such that
\begin{itemize}
    \item $V_{b_\mathrm{R}}/V_{c_\mathrm{R}} > 2^{1-\frac{\mathcal{C}_{\mathrm{l,c}}}{2 z(w)}}$ corresponds to type-I fluctuations, which do not collapse into PBHs;
    \item $1 < V_{b_\mathrm{R}}/V_{c_\mathrm{R}} < 2^{1-\frac{\mathcal{C}_{\mathrm{l,c}}}{2 z(w)}}$ corresponds to type-I fluctuations collapsing into ``type-I PBHs'';
    \item $V_{b_\mathrm{R}}/V_{c_\mathrm{R}} < 1$ delimits the type-II region (type-II fluctuations collapsing into ``type-II PBHs'').
\end{itemize}
From now on we assume that PBHs form in a radiation-dominated universe, where $w=1/3$, and we will adopt the representative value $\mathcal{C}_\mathrm{c} =1/2$ for the type-I threshold, corresponding to $\mathcal{C}_{\mathrm{l,c}}=2/3$. In this case, note that the quantity $2^{1-\mathcal{C}_{\mathrm{l,c}}/\left[2 z(w)\right]}$ merely boils down to $\sqrt{2}$.

We stress that $\mathcal{C}_{\mathrm{c}}=1/2$ should be regarded as a representative choice rather than as a universal collapse threshold. As discussed above, the threshold for type-I perturbations depends on the shape of the compaction-function profile. Since stochastic trees provide only a discrete sampling of the smoothed linear compaction function, the underlying continuous profile, and hence the appropriate profile-dependent threshold, cannot be uniquely reconstructed for each realisation (see discussion around fig.~6). Therefore, the abundance of type-I PBHs is expected to depend quantitatively on the adopted value of $\mathcal{C}_{\mathrm{c}}$.
By contrast, in our prescription, all type-II fluctuations do not involve an additional collapse threshold. The identification of type-II PBHs is therefore not directly affected by the adopted value of the type-I threshold $\mathcal{C}_{\mathrm{c}}$.

\begin{figure}
    \centering
    \includegraphics[width=.8\textwidth]{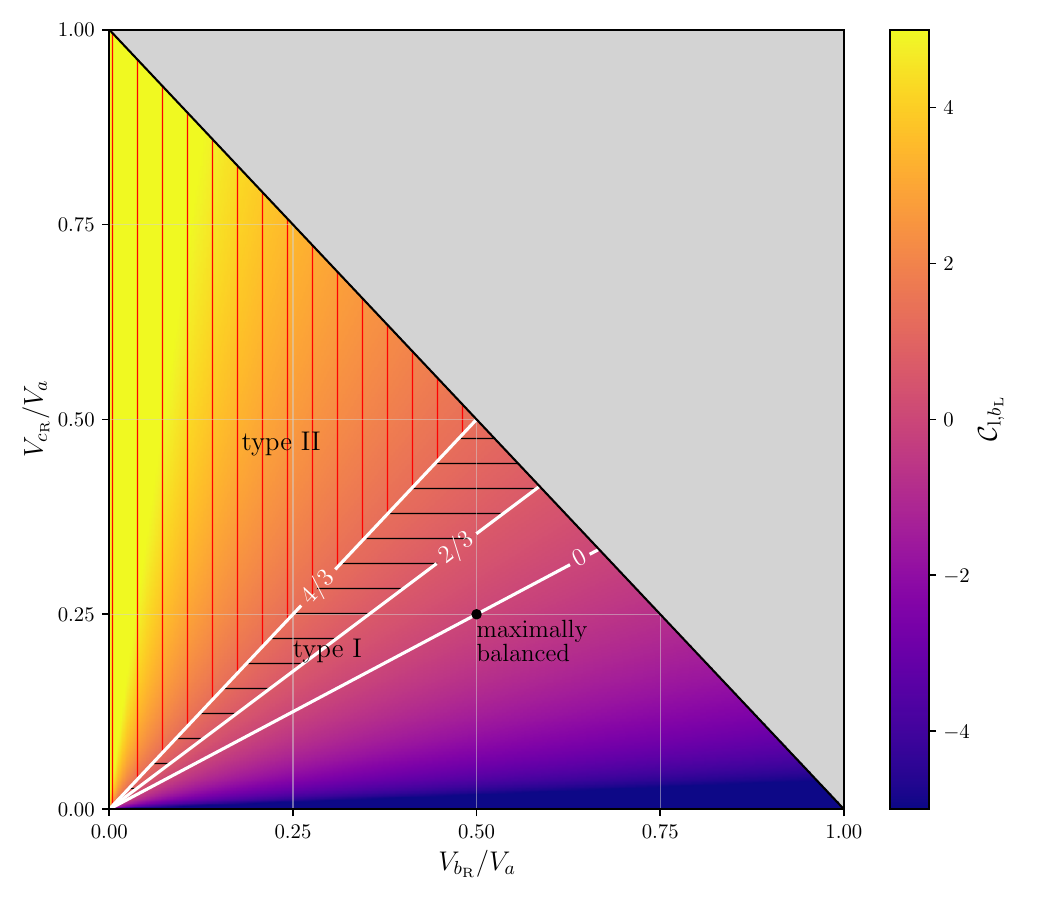}
    \caption{Colour plot of \cref{eq:Cl_offbranch} as a function of $V_{b_\mathrm{R}}/V_a$ and $V_{c_\mathrm{R}}/V_a$, where the domain is restricted according to \cref{eq:constraints:volumes} (the grey masked region lies outside that domain). 
    White contour lines indicate constant values of $\mathcal{C}_{\mathrm{l}, b_\mathrm{L}}=0,2/3,4/3$. The colour scale is clipped to the interval $[-5,5]$ for visualisation purposes, so that the uniformly blue and yellow regions correspond to $\mathcal{C}_{\mathrm{l}, b_\mathrm{L}}<-5$ and  $\mathcal{C}_{\mathrm{l}, b_\mathrm{L}}>5$ respectively. The type-I and type-II PBHs regions are denoted with black and red hatching respectively.
    }
    \label{fig:XY}
\end{figure}

\paragraph{Volume imbalance}

The above criterion implies that if the volume of a node ($V_{c_\mathrm{R}}$) is larger than that of its ``uncle'' ($V_{b_\mathrm{R}}$), then it gives rise to a type-II fluctuation (in the $b_\mathrm{L}$ node), hence producing a type-II PBH. Conversely, if $V_{c_\mathrm{R}}< V_{b_\mathrm{R}}$, we have a type-I fluctuation, which gives rise to a type-I PBH if the volumes are close to each other. All in all, PBH formation results from \emph{volume imbalance}, as mentioned at the end of \cref{sec:compact:trees}.\footnote{This was already highlighted in ref.~\cite{Animali:2025pyf}, albeit with a criterion relying on a ``coarse-shelled'' proxy, see \cref{subsec:comp:coarse:shelled}.} In our setup, the imbalance spreads between the volumes of sibling nodes among the three generations of \cref{fig:nodes_in_derivatives}, quantified by
\begin{equation}
    \frac{V_{b_\mathrm{R}}}{V_a}=1-\frac{V_{b_\mathrm{L}}}{V_a} \qq{and}  \frac{V_{c_\mathrm{R}}}{V_a}=1-\frac{V_{b_\mathrm{R}}}{V_a}-\frac{V_{c_\mathrm{L}}}{V_a}\,,
    \label{eq:X_Y}
\end{equation}
where
\begin{equation}\label{eq:constraints:volumes}
    0\leq \frac{V_{b_\mathrm{R}}}{V_a}\leq 1 \qq{and} 0\leq \frac{V_{c_\mathrm{R}}}{V_a}\leq 1-\frac{V_{b_\mathrm{R}}}{V_a}\,,
\end{equation}
since the volumes are nested. The linear compaction function in \cref{eq:Cl_discrete} is shown in \cref{fig:XY} as a colour plot in terms of these variables. White contour lines indicate constant values; they appear as straight lines in this plane because $\mathcal{C}_{\mathrm{l},b_\mathrm{L}}$ depends only on the ratio $(V_{b_\mathrm{R}}/V_a)/(V_{c_\mathrm{R}}/V_a)$. The highlighted levels are chosen to delineate the boundaries between the three regions discussed above. We see that $\mathcal{C}_{\mathrm{l},b_\mathrm{L}}$ increases toward the upper-left part of the physical domain, corresponding to decreasing $V_{b_\mathrm{R}}/V_a$ and increasing $V_{c_\mathrm{R}}/V_a$. Thus, larger compaction is obtained when less volume is carried by the sibling node $b_{\mathrm{R}}$ and more by the child node $c_{\mathrm{R}}$. The figure further shows that the type-I PBH region (in black hatching) is confined to a relatively narrow band, while the type-II PBH region (in red hatching) occupies a much broader portion of parameter space, indicating that strong imbalance preferentially leads to type-II PBH formation. In the maximally balanced configuration ($V_{b_\mathrm{L}} = V_{b_\mathrm{R}}$, $V_{c_\mathrm{L}}=V_{c_\mathrm{R}}$) we have $V_{b_\mathrm{R}}/V_a = 1/2$ and $V_{c_\mathrm{R}}/V_a = 1/4$. This point is shown as the black dot in the figure, and it lies on the $\mathcal{C}_{\mathrm{l},b_\mathrm{L}}=0$ contour, consistent with the fact that it represents a symmetric tree configuration without volume imbalance and hence with no perturbations.

The maximally imbalanced configuration described later in \cref{subsec:discretization:scheme}, in which the right branch always stops expanding and all right-hand nodes have the same volume $\Delta V$, lies on the diagonal  $V_{b_\mathrm{R}}/V_a = V_{c_\mathrm{R}}/V_a$. Along this line, $\mathcal{C}_{\mathrm{l},b_\mathrm{L}} = 2z(w) = 4/3$ in the radiation-dominated case. In this sense, the line $\mathcal{C}_{\mathrm{l},b_\mathrm{L}} = 4/3$ is precisely the contour associated with this maximally-imbalanced configuration. The corner $(0,0)$ in the colour plot should be interpreted with care. On the above line, it is reached only asymptotically when $\Delta V/V_a \rightarrow 0$, that is, when the constant right volume becomes negligible compared with the parent volume. However, the compaction function is not well-defined in the corner itself, since it depends on the ratio $(V_{b_\mathrm{R}}/V_a)/(V_{c_\mathrm{R}}/V_a)$, which becomes indeterminate there. Moreover, this is not the only case of maximal imbalance: other extreme configurations are located at the other two corners of the physical domain. The point $(1,0)$ corresponds to the limit in which the subtree rooted at $b_{\mathrm{R}}$ contributes all of the final parent volume, and the linear compaction function becomes very negative. Conversely, the point $(0,1)$ corresponds to the limit in which the subtree rooted at $c_{\mathrm{R}}$ dominates the final parent volume and the linear compaction function becomes very positive. In practice, these two limits represent complete transfer of volume to the right branch at the first or second branching, respectively. The fact that the linear compaction function takes different values in these extreme cases shows that it is sensitive not only to the presence of imbalance, but also to the generation in the tree at which that imbalance occurs.

\Cref{fig:numerical_example_tree} depicts an example tree. In accordance with the above discussion, some imbalances around left-hand nodes (indicated by solid arrows) collapse into PBHs. Type-I PBHs (in black) arise from nodes whose right child and right ``uncle'' (indicated by dashed arrows) have similar volumes (both consisting of just one leaf in this example). Type-II PBHs correspond to a greater imbalance, with large right-hand child volumes (consisting of many leaves).

\paragraph{Cloud-in-cloud}

When a subtree deviates from the average trajectory and produces a large volume, it generally induces a substantial degree of imbalance along several generations. 
In other words, when PBH formation becomes significant, black holes tend to ``bloom'' at multiple nodes along a branch, hence they form in a nested manner.\footnote{This did not happen in the example tree of \cref{fig:numerical_example_tree} due to the mildness of the perturbations; see \cref{fig:tree_comparison} for an example of this phenomenon.}
This is precisely the cloud-in-cloud phenomenon~\cite{Jedamzik:1994nr}, in which a hierarchy of overdense regions leads to the formation of PBHs inside larger ones. 
Our stochastic-tree approach, given the recursive nature of $\texttt{FOREST}$ outlined above, automatically accounts for this effect: as soon as the algorithm identifies a PBH at a given node, all PBHs contained within it are discarded~\cite{Animali:2025pyf}. 
Although cloud-in-cloud has been studied in the context of large Gaussian perturbations where almost all PBHs are thought to be of type I~\cite{DeLuca:2020ioi, Auclair:2026kwf}, its physical interpretation for type-II PBHs is still unclear due to the possible fractal nature of these objects. Notwithstanding this concern, we adopt the same procedure to identify and treat both type-I and type-II PBHs.

Notice that this cloud-in-cloud prescription may also introduce an indirect dependence of the final type-I/type-II populations on the adopted type-I collapse threshold. Changing $\mathcal{C}_{\mathrm{ c}}$ may alter which outer regions are identified as PBH forming and, consequently, which inner PBH-forming regions are discarded by the cloud-in-cloud procedure.

\begin{figure}
    \centering
    \includegraphics[width=.32\textwidth]{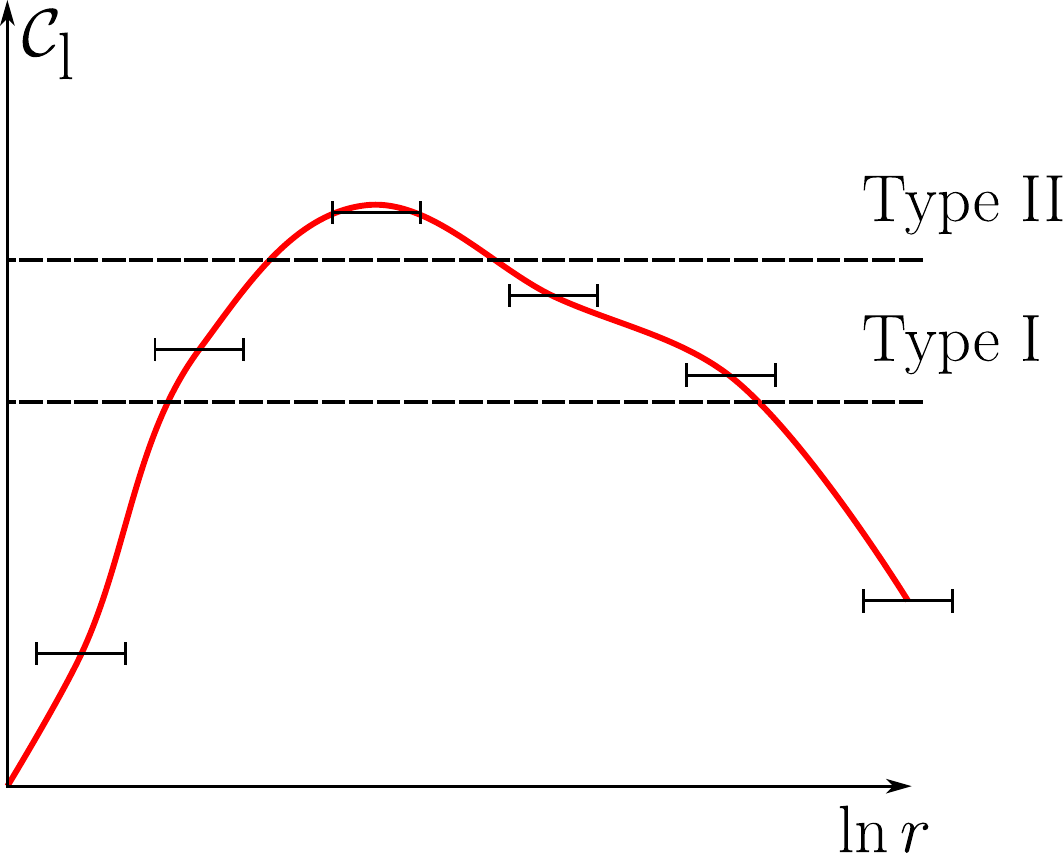}
    \includegraphics[width=.32\textwidth]{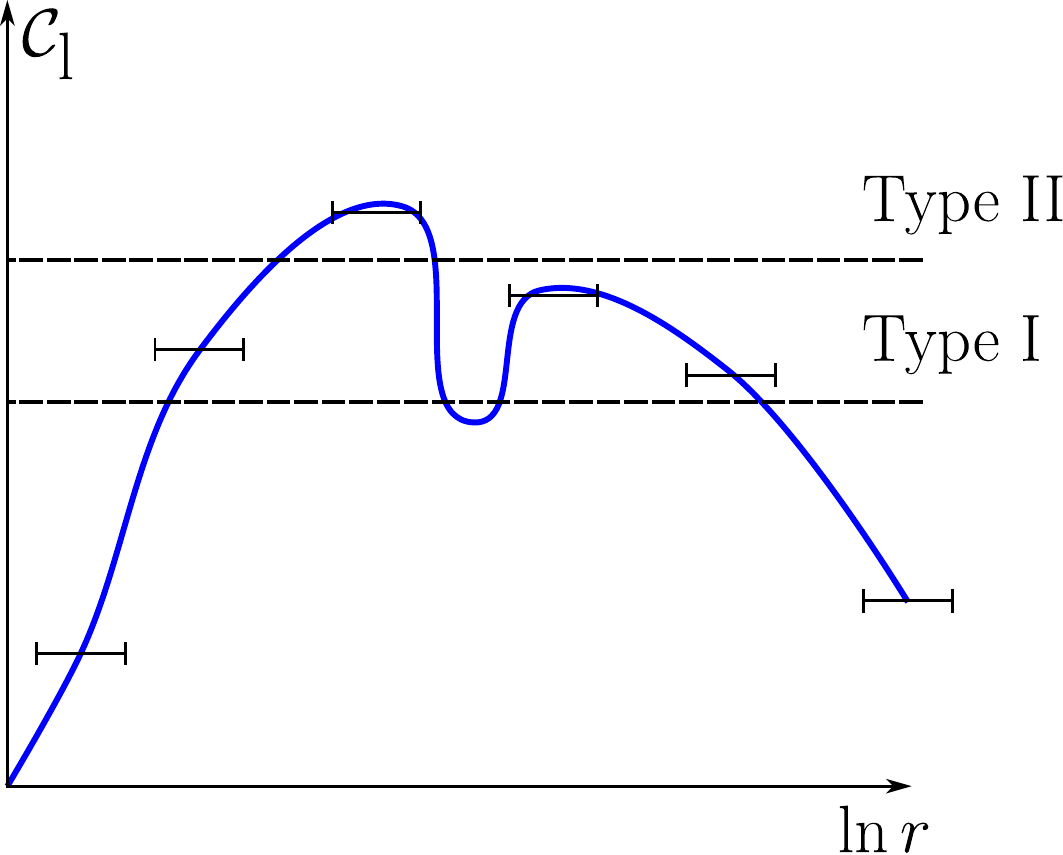}
    \includegraphics[width=.32\textwidth]{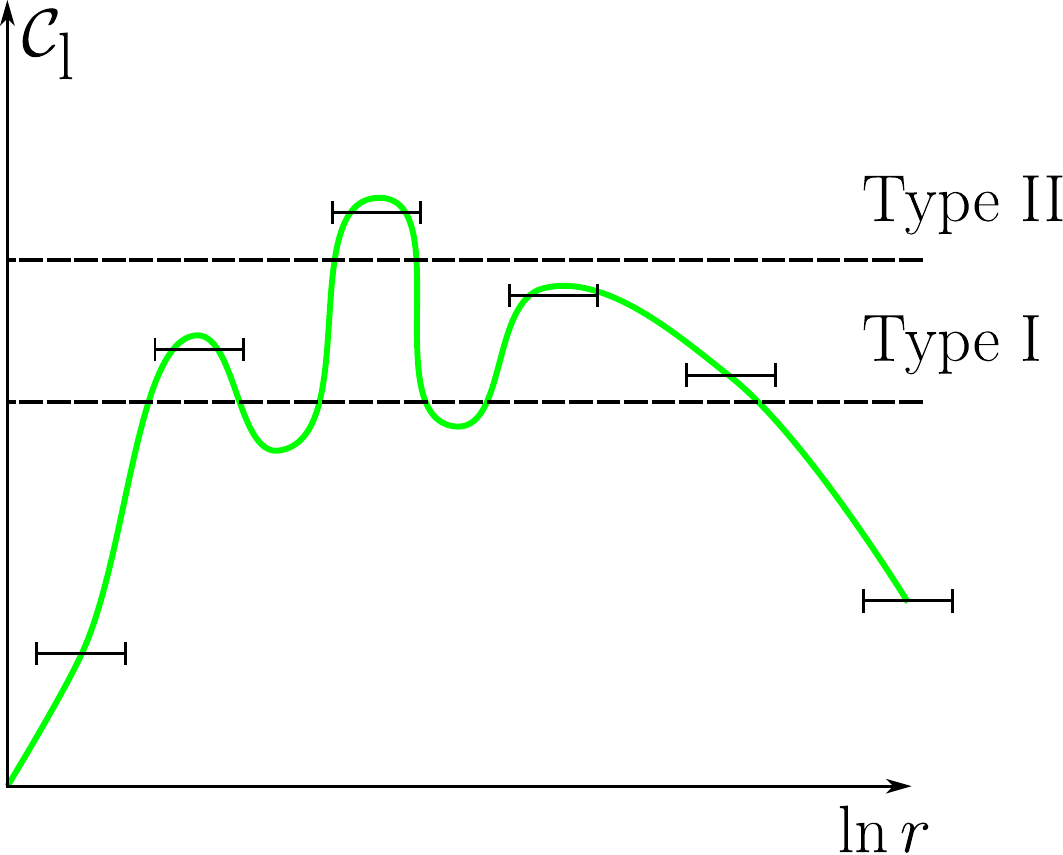}
    \caption{We sample a smoothed version of the linear compaction function over a discrete set of points and we are blind to features at shorter length scales. Here are three possible profiles that provide the same sample of $\mathcal{C}_\mathrm{l}$.
    }
    \label{fig:outermost}
\end{figure}

\paragraph{Limitations}

Let us finally point out that our procedure to identify type-I/II fluctuations is slightly inaccurate. The perturbation type is typically --- for instance in collapse simulations --- determined by the maximum value of $\mathcal{C}_\mathrm{l}$ along the full radial profile, as discussed in \cref{sec:pbh:threshold_I_II}. However, we work here with a smoothed compaction function sampled over a discrete set of points and cannot resolve features shorter than the discretization step. In principle, the actual linear compaction-function profile may exhibit more complicated non-monotonic behaviour along $r$, as illustrated in \cref{fig:outermost}. 

Two main sources of inaccuracy can appear. Firstly, for a smooth profile that first increases and then decreases over a wide range of $r$ values, as displayed in the left panel of \cref{fig:outermost}, our scheme does not retain the linear compaction function at its maximum as normally required, but rather at the largest value of $r$ where it exceeds the collapse threshold. This is why our method may overestimate PBH masses. In addition, even if the linear compaction function lies in the type I region at the largest $r$, it may rise to the type II region further inside. This is why our method may also underestimate the fraction of type II PBHs.

Secondly, for a profile that has one or more narrow maxima, such as in the right panel, the collapse 
dynamics may become more involved.
Our discretised computation may miss high, narrow peaks, leading to an underestimation of the PBH abundance or the type II fraction. The PBH mass may also be underestimated if a high, narrow peak appears at larger $r$. 
Conversely, narrow peaks are associated with high pressure gradients, increasing the collapse threshold \cite{Musco:2018rwt, Escriva:2019phb}, so excluding narrow peaks may actually make the abundance estimate more accurate. Several nearby maxima complicate the situation even more; there is only preliminary work on the fate of such profiles \cite{Nakama:2014fra, Atal:2019erb, Escriva:2023qnq} (of course, if the maxima are well separated, then cloud-in-cloud takes place and our procedure properly captures the effect).

A refinement that could in principle be implemented within stochastic trees would be to consider the chains of consecutive nodes forming primordial black holes, and only retain, in each of these chains, the node presenting the maximum linear compaction function, solving some (though not all) of the above issues. However, if one wants to still take into account cloud-in-cloud, it would also necessitate a criterion on the minimum distance between two chains for them to be considered independent. These issues are anyway all tied to the way the compaction-function theory should be extended beyond the case of moderately-sized, smooth and isolated peaks and framed in presence of large non-Gaussianities, hence the conclusions reached in this work should be taken with some care.

\subsection{Primordial black hole mass}
\label{sec:PBH:mass}

The eventual collapse into a PBH occurs when perturbations re-enter the Hubble radius after inflation.
The mass $M$ of the resulting black hole approximately equals the energy contained within one Hubble volume at that time. The mass depends on the perturbation length scale, that is, on the radius at which $\mathcal{C}$ exceeds the threshold $\mathcal{C}_c$. We remain agnostic about the absolute mass and write
\begin{equation}
    M = M_\sigma \frac{R^2}{R_\sigma^2} \, ,
\label{eq:mass_in_R}
\end{equation}
where $M_\sigma$ is the reference mass of a black hole forming from a comoving region of size $R_\sigma$ on the final hypersurface (the end of inflation or the end of the region with strong quantum diffusion) and $R$ is the PBH length scale. Here we assume instantaneous reheating and radiation domination from the end of inflation until PBH formation. The $R^2$ scaling accounts for the expansion of space and the dilution of the radiation energy density inside the patch in such a scenario \cite{Tomberg:2024chk}. 
Note that, due to the uncertainties mentioned above in the gravitational fate of type-II fluctuations, \cref{eq:mass_in_R} may not apply to them, and the type-II mass distributions we obtain below are indicative only. 

Note that we use the areal radius $R$ rather than the coordinate radius $r$ to determine the black hole mass. For mild perturbations, $\zeta \lesssim 1$, the difference is small; however, for large perturbations, $\zeta \gg 1$, \cref{eq:mass_in_R} ensures that the mass is determined by a physical length scale measured by an observer on the final hypersurface, as it should be.

We neglect critical scaling in the PBH mass near the collapse threshold, i.e.\ when $\mathcal{C} \to \mathcal{C}_\mathrm{c}$ \cite{Musco:2012au}. This typically modifies \cref{eq:mass_in_R} by a factor of order unity.

In our discrete setup, we use $R=R_{a,b}$ from \cref{eq:R_discrete} when $b_\mathrm{L}$ is the outermost PBH-forming node. Note that with this choice, the size of a PBH forming on the left sibling is actually determined by the volume of the right sibling, since $R_{a,b}^3 \propto V_a - V_{b_\mathrm{L}} = V_{b_\mathrm{R}}$, see \cref{eq:R_discrete} and \cref{fig:nodes_in_derivatives}.

\section{Application: tilted quantum-well model}
\label{sec:application}

\subsection{Description of the model}
To illustrate the above framework and provide concrete results, we consider a tilted quantum-well toy model. There the inflationary potential $V(\phi) \equiv 24 \pi^2 \Mp^4 v(\phi) $ features a constant-slope region
\begin{align}
    v(\phi)=v_0 \left(1+\alpha \frac{\phi}{\Mp}\right)
\label{eq:tilted-potential}
\end{align}
between $\phi_{\mathrm{end}}$, where inflation ends (or gives way to a classical phase with negligible quantum diffusion), corresponding to an absorbing boundary, and $\phi_{\mathrm{end}}+\Delta\phi_{\mathrm{well}}$, where the potential is assumed to become steeper and the dynamics of $\phi$ dominated by the classical drift, corresponding to a reflective boundary. For the slow-roll approximation to remain valid, we require $\alpha \ll 1$.

\begin{table}[t]
    \centering
    \begin{tabular}{c c | c | c c c c}
       $d$ & $\mu$ & Classicality $[d\mu^2]$ & $n_\mathrm{trees}$ & $\ev{V} ~ [V_\sigma]$ & $f_{\mathrm{PBH, end}}^{\mathrm{I}}$ & $f_{\mathrm{PBH, end}}^{\mathrm{II}}$ \\ \hline
       \hline
       $0.7$ & $1$ & $0.7$ & $10^{9}$ & $19$ & $0.02$ & $0.16$\\
        $0.7$ & $5$ & $18$ & $10^{9}$ & $485$ & $0.047$ & $0.29$\\
        $0.7$ & $10$ & $70$ & $10^{10}$ & $92$ & $0.26$ & $0.25$\\
        $0.7$ & $20$ & $280$ & $10^{10}$ & $77$ & $0.25$ & $0.027$\\
        $0.7$ & $50$ & $1.8 \cdot 10^{3}$ & $10^{10}$ & $73$ & $0.0051$ & $3.6 \cdot 10^{-8}$\\
        \hline
        $1.0$ & $5$ & $25$ & $10^{10}$ & $27$ & $0.15$ & $0.27$\\
        $1.0$ & $10$ & $100$ & $10^{10}$ & $21$ & $0.21$ & $0.066$\\
        $1.0$ & $20$ & $400$ & $10^{10}$ & $20$ & $0.059$ & $0.00057$\\
        $1.0$ & $50$ & $2.5 \cdot 10^{3}$ & $10^{10}$ & $20$ & $1.8 \cdot 10^{-5}$ & $0$\\
        \hline
        $1.3$ & $1$ & $1.3$ & $10^{9}$ & $5.1$ & $0.042$ & $0.17$\\
        $1.3$ & $1.66$ & $3.7$ & $10^{9}$ & $93$ & $0.0077$ & $0.084$\\
        $1.3$ & $2$ & $5.3$ & $10^{9}$ & $41$ & $0.024$ & $0.17$\\
        $1.3$ & $3$ & $12$ & $10^{9}$ & $14$ & $0.093$ & $0.24$\\
        $1.3$ & $4$ & $21$ & $10^{9}$ & $11$ & $0.13$ & $0.18$\\
        $1.3$ & $5$ & $33$ & $10^{9}$ & $11$ & $0.15$ & $0.12$\\
        $1.3$ & $10$ & $130$ & $10^{9}$ & $9.8$ & $0.098$ & $0.0083$\\
        $1.3$ & $20$ & $520$ & $10^{9}$ & $9.6$ & $0.0066$ & $1.4 \cdot 10^{-6}$\\
        \hline
        $2.0$ & $5$ & $50$ & $10^{10}$ & $4.6$ & $0.063$ & $0.011$\\
        $2.0$ & $10$ & $200$ & $10^{10}$ & $4.5$ & $0.0074$ & $1.2 \cdot 10^{-5}$\\
        $2.0$ & $20$ & $800$ & $10^{10}$ & $4.5$ & $3.8 \cdot 10^{-6}$ & $0$\\
    \end{tabular}
    \caption{Simulations performed in the tilted well. All simulations are initialised with a field value at the reflective boundary $\phi_\mathrm{end} + \Delta \phi_\mathrm{well}$, i.e.\ $x=1$. $d$ and $\mu$ in the first and second columns correspond to the parameters introduced in \cref{eq:d:mu}, the third column displays the classicality parameter $d\mu^2$ discussed in \cref{sec:quantum:classical:critical}, $n_\mathrm{trees}$ in the fourth column corresponds to the number of trees simulated for each $d$-$\mu$ configuration, $\ev{V}$ in the fifth column corresponds to the mean volume obtained from the stochastic-tree population in units of the elementary volume $V_\sigma=4/3 \pi R_\sigma^3$, and finally the last two columns show the integrated mass function of type I and II objects at the end of inflation.
    Note that having $f_{\mathrm{PBH, end}}^{\mathrm{II}} = 0$ means that we have not been able to find a single type II object in the simulation (an upper bound on their abundance can be estimated as $(n_\mathrm{trees} \times \ev{V})^{-1}$ in these cases). 
    }
    \label{tab:sims}
\end{table}

This model, already explored in refs.~\cite{Ezquiaga:2019ftu, Animali:2022otk, Animali:2024jiz, Blachier:2025iwk}, is sufficiently simple to admit accurate analytical approximations in the vacuum-dominated limit where the potential is dominated by its constant term, yet flexible enough to feature a regime dominated by quantum diffusion and one dominated by potential-induced drift, thus exhibiting a well-defined classical limit. The physics is controlled by two dimensionless parameters
\begin{align}
    d =\frac{\alpha \Mp}{\Delta\phi_{\mathrm{well}}}\, ,\qquad 
    \mu^2 = \frac{\Delta \phi_{\mathrm{well}}^2}{v_0 \Mp^2}\, ,
    \label{eq:d:mu}
\end{align}
and it is customary to rescale the inflaton field value according to
\begin{align}
x=\frac{\phi-\phi_{\mathrm{end}}}{\Delta\phi_{\mathrm{well}}}\, ,
\end{align}
such that $0\leq x\leq 1$ within the well. We consider the vacuum-dominated limit $\alpha\Delta\phi_{\mathrm{well}} / \Mp\to 0$, which is taken while keeping $d$ and $\mu$ fixed. Note that this implies letting $\alpha\to 0$ (hence the slow-roll approximation is guaranteed to be satisfied) and $v_0\to 0$ (hence inflation proceeds at sub-Planckian energies and the semi-classical approximation on which stochastic inflation rests also holds), so this limit is physically relevant. The Langevin equation~\eqref{eq:Langevin} reduces to
\begin{align}
\label{eq:Langevin:tilted:well}
 \frac{\dd x}{\dd N} =  -d+\frac{\sqrt{2}}{\mu} \xi(N) \, ,
\end{align}
and describes Brownian motion with constant drift $d$ and diffusion amplitude $\sqrt{2}/\mu$.

\begin{figure}
    \centering
    \begin{subfigure}[t]{.49\textwidth}
		\includegraphics[width=\textwidth]{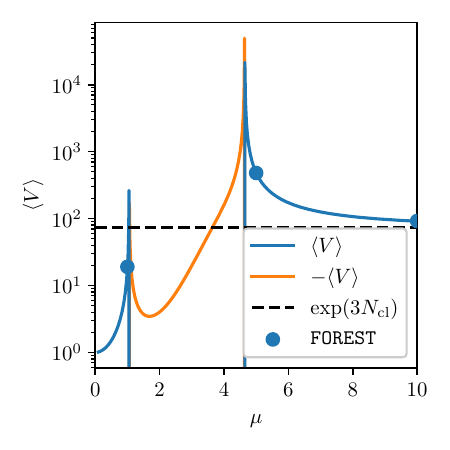}
		\caption{$d = 0.7$}
	\end{subfigure}
	\begin{subfigure}[t]{.49\textwidth}
		\includegraphics[width=\textwidth]{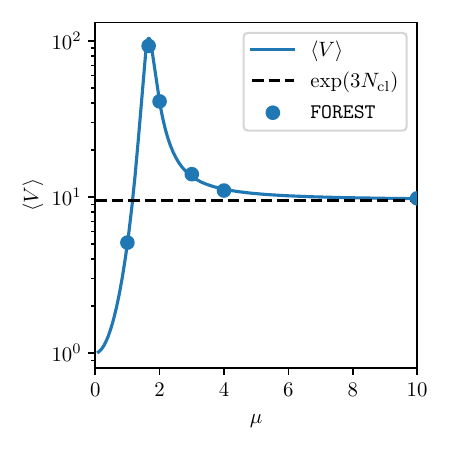}
		\caption{$d = 1.33$}
	\end{subfigure}
    \caption{Mean volume for an initial field value at the reflective boundary $x=1$ and for different values of $d$, as a function of $\mu$ in the tilted quantum well, using the characteristic function~\eqref{eq:char:funct:tilted} evaluated at $t=-3i$. The black dashed line represents the classical expectation $e^{3 N_{\mathrm{cl}}}$, where $N_{\mathrm{cl}}=1/d$ is the (deterministic) number of \efolds realized in the absence of diffusion, and which is reached asymptotically in the classical limit $d\mu^2\rightarrow \infty$. The blue dots show numerical results obtained with \texttt{FOREST}, and are in excellent agreement with the analytical prediction. 
    }
    \label{fig:volume}
\end{figure}

\subsection{Quantum, classical and critical regimes}\label{sec:quantum:classical:critical}

\begin{figure}
    \centering
    \includegraphics[width=0.9\textwidth]{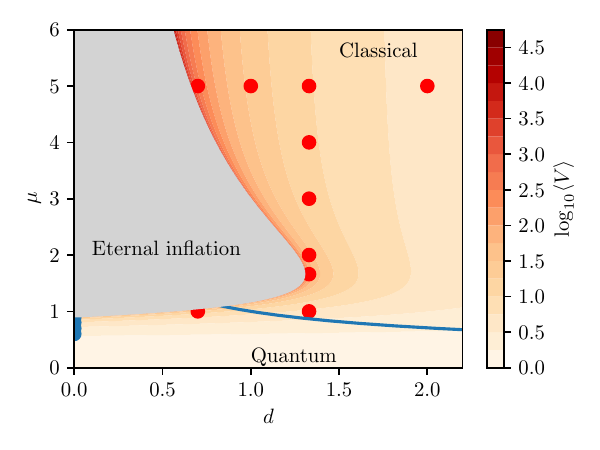}
    \caption{Mean tree volume in the tilted quantum well, in the $(d, \mu)$ parameter space and starting from the reflective boundary. 
    The region in grey leads to ``eternal inflation'', with $\ev{V} \to \infty$.
    $\ev{V}$ increases rapidly at the boundary of this region (the ``critical boundary'').
    At low $\mu$, the system is dominated by quantum diffusion and $\ev{V} \to 1$.
    If the drift $d$ is non-zero, there exists a value for $\mu$ above which 
    the classical drift dominates and $\ev{V} \approx e^{3 / d}$ as $\mu \to \infty$.
    The blue line displays $d\mu^2 = 1$.
    Blue points represent simulations of the flat quantum well ($d=0$) performed in~\cite{Animali:2025pyf}, while red points correspond to the tilted case ($d\neq 0$) simulations presented here and listed in \cref{tab:sims}. 
    }
    \label{fig:phase_diagram}
\end{figure}

In the absence of stochastic noise in \cref{eq:Langevin:tilted:well}, the classical number of \efolds $N_{\mathrm{cl}}=x/d$ starting from field value $x$ is realized. Over that time, the typical stochastic spread scales as $\mu^{-1}\sqrt{N_{\mathrm{cl}}}\propto (\mu^2 d)^{-1/2}$. This defines a classicality parameter, which controls how far from the classical limit the system lies. For instance, if $\mathcal{N}$ denotes the first-passage time through the absorbing boundary located at $\phi_{\mathrm{end}}$, its mean value starting from field value $x$ reads~\cite{Animali:2024jiz}
\begin{align}
\left\langle\mathcal{N}\right\rangle = \frac{x}{d} + e^{-d\mu^2}\frac{1-e^{d\mu^2 x }}{d^2\mu^2}\, .
\end{align}
When $d\mu^2\gg 1$, the first term on the right-hand side dominates, which coincides with the classical number of \efolds $N_{\mathrm{cl}}$. In what follows, we will thus refer to
\begin{itemize}
    \item $d \mu^2 \gg 1$ as the classical regime,
    \item $d \mu^2 \ll 1$ as the quantum regime.
\end{itemize}
One should however stress that, on the (exponential) tail of the first-passage-time distribution, the (Gaussian) classical limit always breaks down~\cite{Ezquiaga:2019ftu}. This is why the above nomenclature is not a statement about the prominence of quantum-diffusion effects on PBHs, which form in tails, but rather characterizes the bulk of the first-passage-time distribution. We will nonetheless use it to organize parameter-space exploration.

The interplay between the classical and quantum regimes can also be studied at the level of the mean volume $ \langle V \rangle$ originating from a patch with given initial field value $x$. In practice, this quantity can either be computed analytically from the characteristic function of the first-passage-time distribution and using that $ \langle V \rangle = \langle e^{3 \mathcal{N}} \rangle$, see \cref{subsec:vol:weigh:FPT}, or numerically by simulating a large population of trees using \texttt{FOREST}. As shown in \cref{fig:volume}, both methods yield consistent results. 

In the figure, $\langle V \rangle$ is plotted against $\mu$ for two distinct values of $d$. When $d$ is smaller than a critical value $d_\mathrm{c} \sim 1.3$, the mean volume is not well-defined for all $\mu$; as can be seen in panel (a), there exists a region where $\langle V \rangle$ as given from the characteristic function takes unphysical negative values. These negative values indicate divergences in the mean volume and are associated with the appearance of eternal inflation. This pathological region is nonetheless bounded: for a fixed value of $d$, the mean volume remains positive and convergent when $\mu$ is either sufficiently small or sufficiently large. For $d>d_{\mathrm{c}}$, the mean volume is well-defined and convergent for all values of $\mu$. As shown in panel (b) of \cref{fig:volume}, it reaches a maximum for a value of $\mu$ such that $d \mu^2$ is of order unity, a regime where drift and diffusion contribute comparably to the dynamics.

\begin{figure}
	\begin{subfigure}{.49\textwidth}
		\includegraphics[width=\textwidth]{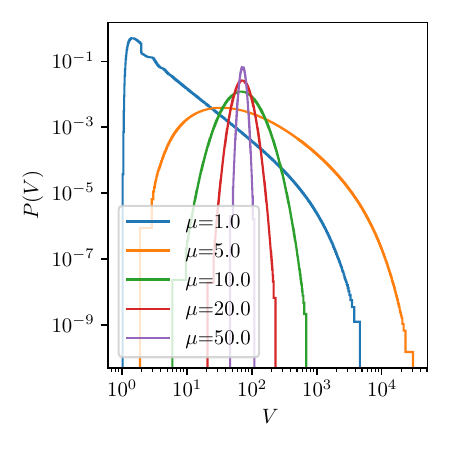}
		\caption{$d = 0.7$}
	\end{subfigure}
	\begin{subfigure}{.49\textwidth}
		\includegraphics[width=\textwidth]{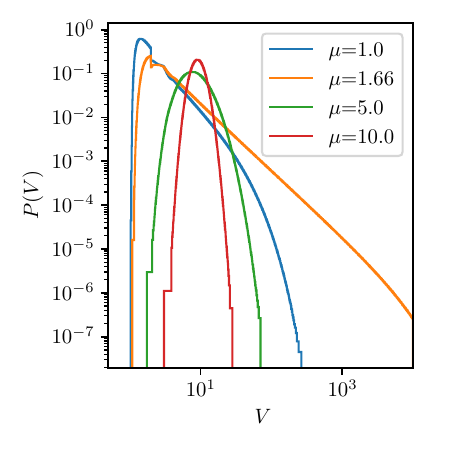}
		\caption{$d = 1.33$}
	\end{subfigure}
	\caption{Equal-density histograms for the probability distribution of the final volume $V$ over tree populations in the tilted well for different values of $d$ and $\mu$. These simulations are listed in \cref{tab:sims}. Each bin contains $10^4$ samples.}
	\label{fig:vol-tree}
\end{figure}

These observations can be generalized by plotting $\langle V \rangle$ over the $(d,\mu)$ parameter space, as shown in \cref{fig:phase_diagram}. The two panels in \cref{fig:volume} correspond to slices of this diagram at constant $d$. The $d=0$ slice corresponds to the flat quantum-well case explored in the previous work~\cite{Animali:2025pyf} (whose results are indicated by blue dots in \cref{fig:phase_diagram}). The phase diagram thus illustrates how the extension to the tilted-well potential allows for the exploration of a much richer phenomenology.

In \cref{fig:phase_diagram}, the grey region delineates the ``eternal inflation'' regime, leading to divergent mean volumes. In the quantum regime, diffusion quickly kicks the field out of the well and the mean volume remains low. As $d\mu^2\to 0$, $\langle V\rangle \to 1$, corresponding to trees with a single node. In the classical regime $d\mu^2\gg 1$,  $\langle V\rangle \to e^{3/d}$ and larger trees are produced. Note however that this limit is only reached asymptotically, and much larger volumes can be produced if the system lies close enough to criticality (i.e.\ close enough to eternal inflation). This is due to the presence of heavy tails in that region, which imply that trees much larger than the classical expectation can be produced even when $d\mu^2$ is large.

The role of tails near criticality is confirmed in \cref{fig:vol-tree}, where the probability distribution of the volume, $P(V)$, is displayed.
Contrary to Ref. \cite{Animali:2025pyf}, in which results were presented in \emph{equal-width} histograms, we changed our binning strategy to have \emph{equal-density} histograms.
In practice, each bin in the histogram is enlarged until it contains at least $10^4$ samples. 
Therefore, we expect the error on the amplitude of each bin to be of the order of the percent by virtue of the central-limit theorem.
As a consequence, the statistical uncertainties are manifested by larger bin widths at low abundances.

In the classical regime the distribution is quasi Gaussian and becomes increasingly peaked around its mean as $d\mu^2$ increases. In contrast, the distribution develops heavier tails in the quantum regime, characterized by power-law decay followed by exponential falloff~\cite{Animali:2025pyf}. Near criticality, that exponential cut occurs deeper in the tail, and eventually disappears in the eternal-inflation region, which explains why the mean volume is divergent there.

Finally, note that in the flat-well case ($d=0$) considered in ref.~\cite{Animali:2025pyf}, small volumes are produced unless parameters are chosen close to criticality. This is why considering the tilted well is necessary to probe beyond the near-critical zone.

\begin{figure}[t!]
	\begin{subfigure}{.49\textwidth}
		\includegraphics[width=\textwidth]{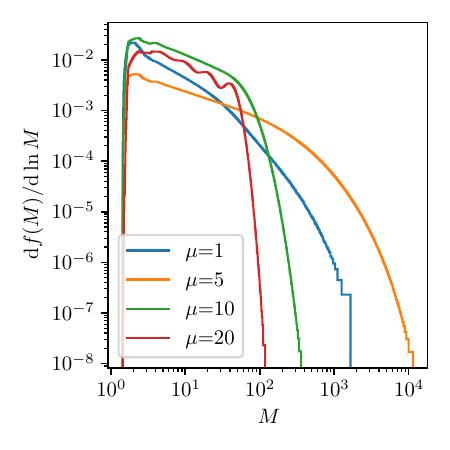}
		\caption{$d = 0.7$}
	\end{subfigure}
	\begin{subfigure}{.49\textwidth}
		\includegraphics[width=\textwidth]{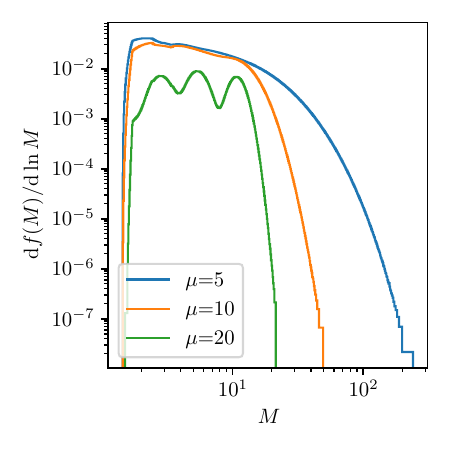}
		\caption{$d = 1.0$}
	\end{subfigure}
	\begin{subfigure}{.49\textwidth}
		\includegraphics[width=\textwidth]{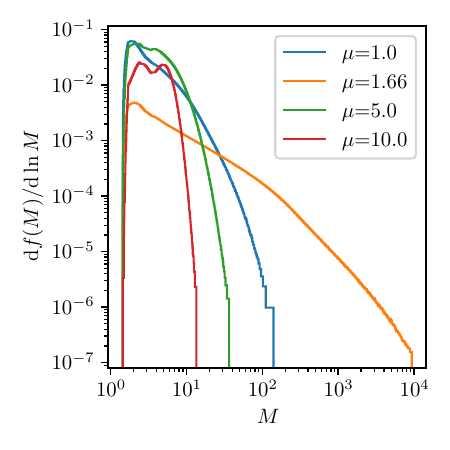}
		\caption{$d = 1.33$}
	\end{subfigure}
	\begin{subfigure}{.49\textwidth}
		\includegraphics[width=\textwidth]{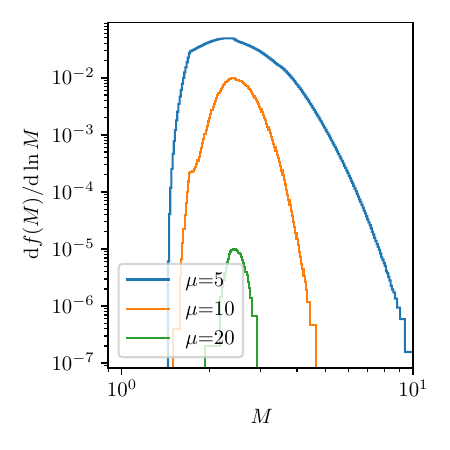}
		\caption{$d = 2.0$}
	\end{subfigure}
	\caption{Equal-density histograms for the distribution of masses for the PBHs produced in the tilted-well model for different values of $\mu$ and $d$.
	Simulation details are listed in \cref{tab:sims}. While the mass distribution remains narrow in the classical regime (panel (d)), it extends over several orders of magnitude when quantum diffusion becomes substantial.}
	\label{fig:pbh-tree}
\end{figure}

\subsection{Distribution of PBH populations}
\label{subsec:pbhs:distr}

As described in detail in \cref{sec:BH:stoch:trees}, \texttt{FOREST} identifies and records all PBHs that appear in our populations of stochastic trees by implementing the sampled version of the compaction function introduced in \cref{sec:compact:trees}. As each tree is explored, the algorithm retains only the outermost, and therefore largest PBHs within each chain of consecutive collapsing nodes. We then assume that the entire downstream branch collapses, which sets the mass of the resulting PBH, see \cref{sec:PBH:mass}.

\Cref{fig:pbh-tree} shows the resulting PBHs mass distribution, irrespective of whether the PBHs are of type I or type II, for different values of the drift $d$ and the diffusion parameter $\mu$. The quantity $\dv*{f(M)}{\ln M} \times \dd{\ln M}$ is defined as the fraction of the universe, evaluated at the end of inflation (or at the end of the tilted well), that will eventually collapse into PBHs with masses comprised between $\ln M$ and $\ln M+\dd\ln M$.\footnote{Note that $f(M)$, as defined here, does not take into account the mass-dependent dilution of the PBH density after the collapse. It describes perturbation statistics at the end of inflation, not PBH abundance or mass distribution today.} The integrated mass fraction reported in \cref{tab:sims} is given by $f_\text{PBH,end} = \int \dv*{f(M)}{\ln M} \times \dd{\ln M}$.

In the near-critical regime, the PBH mass function extends over several orders of magnitude. This is because anomalously large trees are produced in this case, which give rise to large black holes. The behaviour is similar to that of the volume distribution, see \cref{fig:vol-tree}. By contrast, in the classical regime --- panel (d) --- the mass function is narrower and extends over $\Delta\ln M \sim 2 N_{\mathrm{cl}}=2/d$. Let us highlight that even in this classical limit, ``cloud-in-cloud'' still affects the resulting mass function since a suppression of low-mass tails is noticeable.

Note also that in panels (a) and (b), deep in the classical limit, the mass function features unphysical oscillations. This is because, in this regime, most trees are almost perfectly balanced, hence subtrees come with volumes that can only be powers of two (in units of $V_\sigma$). This is obviously a discretization artefact: our scheme cannot be used to probe the details of the mass function beyond the discretization scale, and when smoothing over that scale, such oscillations disappear.

\begin{figure}
	\begin{subfigure}{.325\textwidth}
		\includegraphics[width=\textwidth]{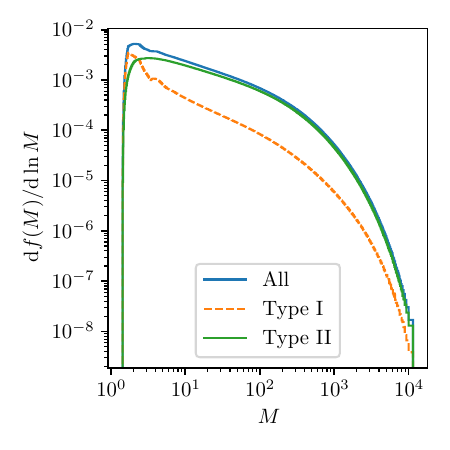}
		\caption{$d = 0.7, \mu=5$}
		
		\includegraphics[width=\textwidth]{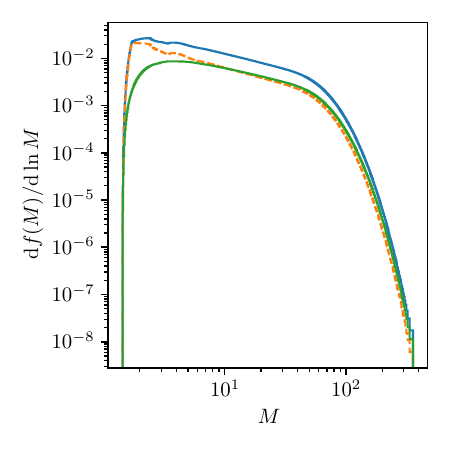}
		\caption{$d = 0.7, \mu=10$}
	\end{subfigure}
	\begin{subfigure}{.325\textwidth}
		\includegraphics[width=\textwidth]{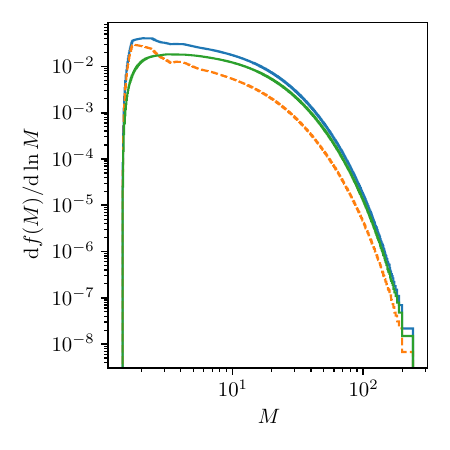}
		\caption{$d = 1.0, \mu=5$}
		
		\includegraphics[width=\textwidth]{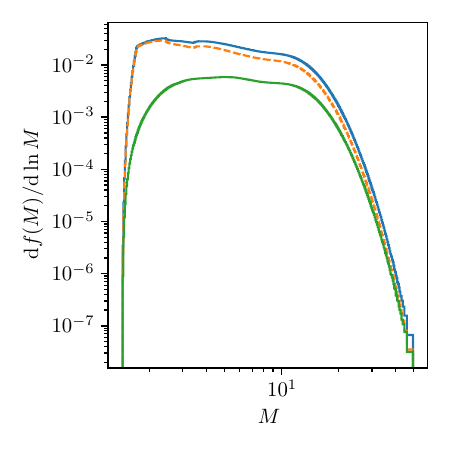}
	    \caption{$d = 1.0, \mu=10$}
	\end{subfigure}
	\begin{subfigure}{.325\textwidth}
		\includegraphics[width=\linewidth]{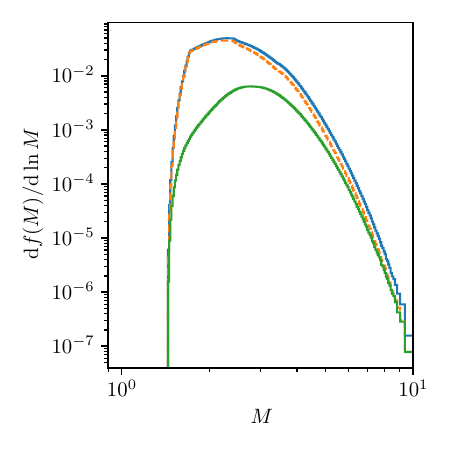}
		\caption{$d = 2.0, \mu=5$}

		\includegraphics[width=\linewidth]{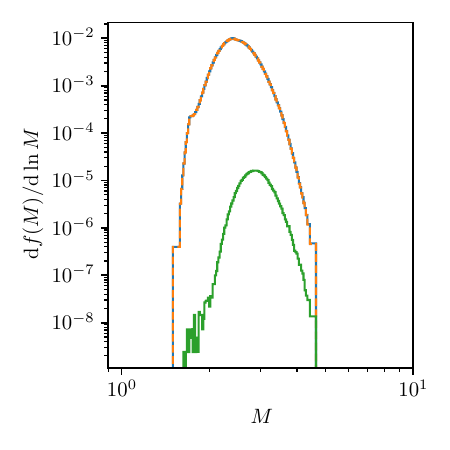}
		\caption{$d = 2.0, \mu=10$}
	\end{subfigure}
	
	\caption{Equal-density histograms for the type-I and type-II PBH mass functions, compared to the overall PBH mass function, for different values of $\mu$ and $d$ in the tilted well. Simulation details are listed in \cref{tab:sims}.
    }
	\label{fig:pbh-tree-1-and-2}
\end{figure}

\subsection{Type-I versus type-II PBHs}

Let us now refine the analysis and distinguish type-I and type-II PBHs. \Cref{fig:pbh-tree-1-and-2} shows the overall PBH mass distribution, together with the separate contributions from type-I and type-II PBHs, identified according to the criteria detailed in \cref{sec:BH:stoch:trees}. We recall that, as outlined in \cref{sec:pbh:threshold_I_II}, by convention we call type-II ``PBH'' any object arising from the gravitational evolution of a type-II fluctuation, regardless of the actual nature of the collapsed object formed from such configurations.

In the classical regime, see especially panels (e) and (f), type-II PBHs are rare and most PBHs are of type I (see also the integrated mass fractions in \cref{tab:sims}). This is consistent with the standard expectation that type-II PBHs require larger overdensities than type-I PBHs to be produced, and are therefore exponentially suppressed when the fluctuations are nearly Gaussian. Note that, as recently put forward in ref.~\cite{Fumagalli:2024kgg}, this holds for narrow power spectra, which is the case in our nearly Gaussian limit due to how peaked the resulting mass distributions are; otherwise, bimodal patterns may appear.

Closer to the quantum or near-critical regimes, the type-II population is less suppressed and the degeneracy between the overall abundance and the type-I contribution is progressively lifted. Type-II PBHs may even account for nearly the entire PBH abundance, which confirms the assumption of refs.~\cite{Gow:2022jfb, Escriva:2023uko} and shows that strong non-Gaussianities, beyond the logarithmic or local-$f_{\mathrm{NL}}$ ansatzes~\cite{Shimada:2024eec, Inui:2024fgk, Escriva:2025eqc}, can dramatically enhance the production of type-II perturbations.

We stress that the precise relative abundance of type-I and type-II objects is expected to depend quantitatively on the adopted type-I collapse threshold. However, the qualitative trend found here that type-II fluctuations are progressively enhanced as quantum-diffusion effects become important is not tied to the specific adopted choice $\mathcal{C}_{\mathrm{c}}=1/2$.

In terms of shape, whenever the type-II contribution is non-negligible, its mass function appears to be systematically broader and less peaked than the corresponding type-I one. In particular, when the distributions exhibit a power-law followed by an exponential tail, the power-law slope is shallower for type-II PBHs than for type-I PBHs. One should note that while the exponential tail is a signature of stochastic effects, the mild power-law behaviour at large PBH masses can be interpreted as arising from ``cloud-in-cloud'' effects~\cite{Animali:2025pyf}. As already discussed in \cref{sec:BH:stoch:trees}, however, the structure of spacetime may be highly intricate when type-II PBHs become abundant and in this regime it is not clear whether the notion of ``cloud-in-cloud'' is still relevant. Furthermore, as highlighted in \cref{sec:PBH:mass}, the prescription used to compute the mass of the PBHs might fail for objects formed from type-II configurations. Therefore, the specific shape of the type-II mass function should be interpreted with caution. 

All in all, whenever stochastic effects become dominant, a large abundance of type-II perturbations seems to be an inevitable outcome, which raises important questions on the fundamental modelling of the objects resulting from their collapse.

\section{Discussion}
\label{sec:discussion}

\subsection{Discretization scheme}
\label{subsec:discretization:scheme}

\begin{figure}
    \centering
    \includegraphics[width=\textwidth]{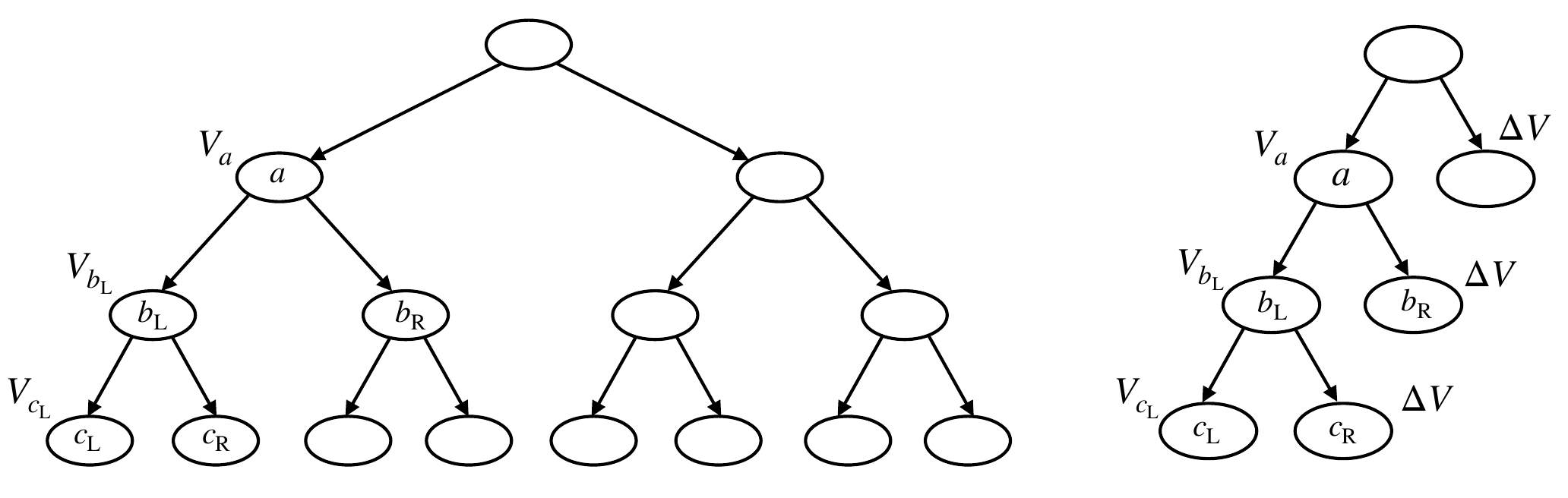}
    \caption{Maximally balanced (left) and maximally imbalanced (right) stochastic-tree configurations.}
    \label{fig:limiting_cases_trees}
\end{figure}
Stochastic trees are discrete in both space and time, and the quantities required to characterize the compaction function (here, the volume enclosed within a given coordinate radius $r$) can only be accessed at the nodes. This motivates the discretization scheme introduced in \cref{sec:compact:trees}. A desirable scheme should preserve the key properties of the corresponding continuum limit, \cref{eq:R3,eq:C_in_V}; let us show this is indeed the case for \cref{eq:R_discrete,eq:Cl_discrete}. In particular, \cref{eq:Cl_discrete}, the discretized version of the linear compaction function $\mathcal{C}_{\mathrm{l},b_{\mathrm{L}}}$, is such that:
\begin{itemize}
    \item it remains invariant under translations $V_j \to V_j + \Delta V$ ($\Delta V=\text{const.}$) along the main chain of nodes ($j=a,b_{\mathrm{L}},c_{\mathrm{L}}$). Such a shift corresponds to adding an extra volume contribution at some deeper point along the chain. Since the compaction function is a local quantity, it should be insensitive to such an insertion;
    \item it remains unchanged under uniform volume rescaling, namely when $V_j \to c V_j$ ($c=\text{const.}$) for all nodes. Such a rescaling corresponds to a renormalization of the metric scale factor and should therefore leave physical quantities, such as the compaction function, unaffected. Physically, this may be interpreted as an overall homogeneous rescaling of the expansion history, for example due to an additional phase of semiclassical inflation after the tilted quantum-well regime, or to a prolonged reheating epoch;
    \item it crosses the critical value $2z(w)$ when the areal radius no longer grows monotonically with the radial distance, see \cref{footnote:typeII:criterion:discretized}.
\end{itemize}
We also note that the discretized expression does not impose any artificial restriction on the range of the linear compaction function, which can take arbitrary values. 

Moreover, the discretization scheme yields the expected physical behaviour of the compaction function in some limiting cases of interest, which we discuss now. In a maximally balanced tree, inflation ends simultaneously in all branches, so the tree is maximally symmetric. This is a generalization of the maximally balanced subtrees discussed in \cref{sec:compact:trees} and is depicted schematically on the left panel in \cref{fig:limiting_cases_trees}. As a result, the curvature perturbation vanishes, and so does the compaction function. For any chain of three consecutive nodes one has $V_a =2 V_{b_\mathrm{L}}=4 V_{c_\mathrm{L}}$, $V_a - V_{b_\mathrm{L}} = 2 V_{c_\mathrm{L}}, V_{b_\mathrm{L}}-V_{c_\mathrm{L}} = V_{c_\mathrm{L}}$, and substituting these relations into \cref{eq:R_discrete,eq:Cl_discrete} yields $R_{a,b}^3 = 3/(2 \pi \ln(2)) V_{c_\mathrm{L}}=2 R_{b,c}^3$, so that $R^3$ grows by a factor 2 between nodes, and $\mathcal{C}_{\mathrm{l}, b_\mathrm{L}}=0$.

Perturbations break this maximally symmetric configuration and induce an imbalance in the tree structure. An extreme example is obtained when, along a given chain of nodes, the right child always stops expanding immediately after each branching event, and therefore remains at volume $\Delta V= V_\sigma/2$, while the left child continues to grow. When this pattern is repeated recursively until the end-of-inflation condition is reached also along the left branch, we refer to this limiting configuration as the maximally imbalanced case. It is depicted in the right panel in \cref{fig:limiting_cases_trees}. There, the compaction function is expected to be maximal, since $\dd V/\dd \ln r$ is a constant, and thus $\mathcal{C}_\text{l} = 2z(w)$ according to the continuum equation \eqref{eq:C_in_V}. In the discrete setup, for any three consecutive nodes along the main chain, one has $V_a = V_{b_\mathrm{L}}+\Delta V = V_{c_\mathrm{L}}+2 \Delta V$, so that $V_a - V_{b_\mathrm{L}} = V_{b_\mathrm{L}} - V_{c_\mathrm{L}} = \Delta V$, where $\Delta V$ is given above. In this case, \cref{eq:R_discrete,eq:Cl_discrete} yield
\begin{equation}
    R_{a,b}^{3}=R_{b,c}^{3} = \frac{3}{4 \pi \ln(2)} \Delta V \qq{and} \mathcal{C}_{\mathrm{l},b_\mathrm{L}} = 2 z(w) \, ,
\end{equation}
as they should.
Since the change in $V$ is constant, $R$ also remains constant along the chain and the compaction function saturates at its maximal value. Note that this is also true for other, non-minimal right-hand volumes $\Delta V$, as long as they stay equal along the chain.

\subsection{Volume weighted first-passage-time distribution}
\label{subsec:vol:weigh:FPT}

\begin{figure}
	\begin{subfigure}[t]{.49\textwidth}
		\includegraphics[width=\textwidth]{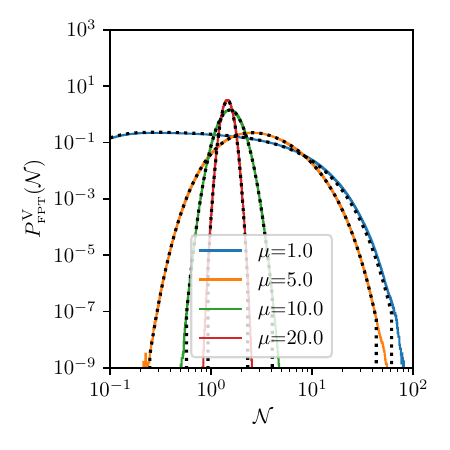}
		\caption{$d = 0.7$}
	\end{subfigure}
	\begin{subfigure}[t]{.49\textwidth}
		\includegraphics[width=\textwidth]{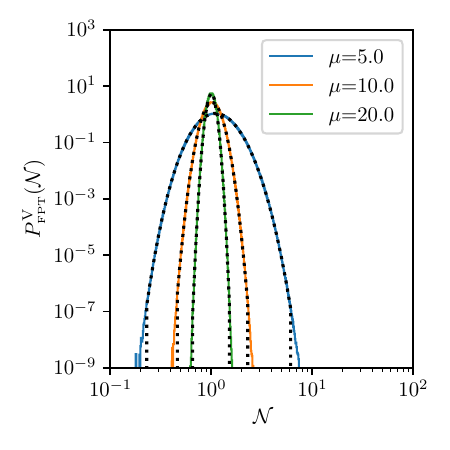}
		\caption{$d = 1.0$}
	\end{subfigure}
	\begin{subfigure}[t]{.49\textwidth}
		\includegraphics[width=\textwidth]{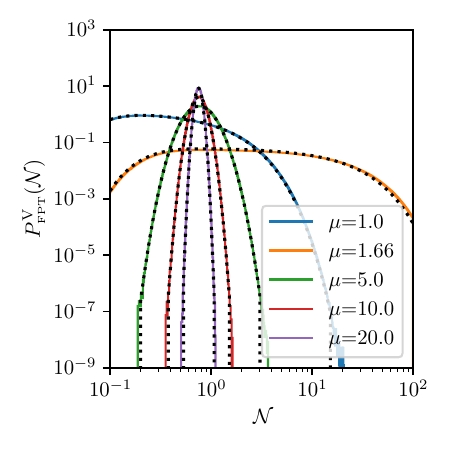}
		\caption{$d = 1.33$}
	\end{subfigure}
	\begin{subfigure}[t]{.49\textwidth}
		\includegraphics[width=\textwidth]{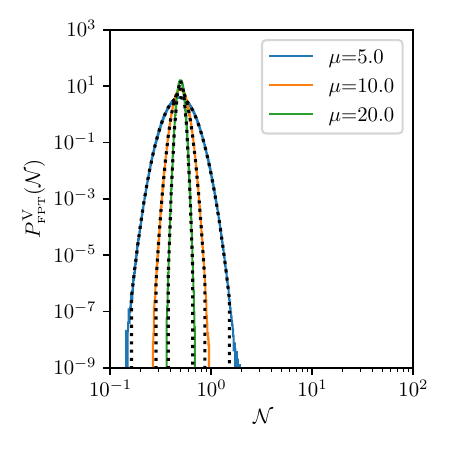}
		\caption{$d = 2.0$}
	\end{subfigure}
	\caption{Volume weighted first-passage-time distribution in the tilted well, for $d=0.7, 1.0, 1.33, 2.0$ and a few values of $\mu$. The initial condition is set to the reflective boundary $x=1$. Solid lines are obtained from the stochastic-tree simulations listed in \cref{tab:sims}. 	Overlapping black-dotted lines correspond to the semi-analytical result obtained by inverse Fourier transforming the volume-weighted characteristic function in \cref{eq:char:funct:tilted}. 
	}
	\label{fig:zeta-tree}
\end{figure}

In addition to the statistics presented in \cref{sec:application}, reconstructed over the set of trees, we can also display statistics over the leaves of the entire tree population, i.e.\ over the forest. For instance, given a tree realization, one can record for each leaf the number of \efolds elapsed from the root node and weight it by the leaf's final volume, yielding the volume-weighted first-passage-time distribution through the end-of-inflation hypersurface.

The latter can be compared with the analytical prescription
\begin{equation}
\PfptV{\phi}(\mathcal{N})=\frac{\Pfpt{\phi}(\mathcal{N})e^{3\mathcal{N}}}{\int_0^\infty \Pfpt{\phi}(\mathcal{N})e^{3\mathcal{N}} \dd \mathcal{N}}\,,
\end{equation}
which can be extracted in a semi-analytical way by inverse Fourier transforming the characteristic function~\cite{Animali:2024jiz}
\begin{equation}
\chi^{\mathrm{V}}_{\mathcal{N}}(t,\phi)=\int_{-\infty}^\infty \dd \mathcal{N} e^{i t \mathcal{N}}\PfptV{\phi}(\mathcal{N})
=\frac{\chi_{\mathcal{N}}(t-3 i,\phi)}{\chi_{\mathcal{N}}(-3 i,\phi)}\, ,
\end{equation}
where
\begin{equation}
\label{eq:char:funct:tilted}
\begin{split}
\chi_{\mathcal{N}}(t,\phi)&=\int_{-\infty}^\infty \dd \mathcal{N} e^{i t \mathcal{N}}\Pfpt{\phi}(\mathcal{N})\\
&=e^{\frac{d  \mu^2 x}{2}} \frac{ \sqrt{4 i t - d^2 \mu^2} \cos{\left(\frac{x-1}{2}\sqrt{4 i t- d^2 \mu^2} \mu\right)}- d \mu \sin{\left(\frac{x-1}{2} \sqrt{4 i t- d^2 \mu^2} \mu\right)}}{ \sqrt{4 i t - d^2 \mu^2}\cos{\left(\frac{1}{2} \sqrt{4 i t - d^2 \mu^2}\mu\right)+d \mu \sin{\left(\frac{1}{2} \sqrt{4 i t - d^2 \mu^2}\mu\right)} }}
\end{split}
\end{equation}
is the characteristic function of the unweighted first-passage-time distribution in the tilted well.

The semi-analytical distribution is displayed in \cref{fig:zeta-tree}, together with the distributions obtained from stochastic-tree simulations, for different values of $\mu$ and $d$.
We observe excellent agreement between the two, for all values of $d$ and $\mu$ considered and for any value of $\mathcal{N}$. This provides a robust validation of the numerical implementation, including the convergence of the solver for the stochastic differential equation governing the field dynamics. Moreover, it offers a non-trivial consistency check of the volume-weighting procedure introduced in refs.~\cite{Tada:2021zzj, Animali:2024jiz} and already verified in the flat-well scenario~\cite{Animali:2025pyf}.

The behaviour of the distribution is in line with expectations: in the classical limit, characterized by large values of $d \mu^2$, it approaches a Gaussian shape, whereas progressively stronger deviations from Gaussianity appear as $d \mu^2$ decreases. In the quantum and near-critical regimes, the distribution features a power-law region followed by an exponential tail, most clearly seen for $d=0.7, \mu=5$ in panel (a) of \cref{fig:zeta-tree}.

\subsection{Comparison with the coarse-shelled curvature perturbation proxy}
\label{subsec:comp:coarse:shelled}

\begin{figure}
    \centering
    \includegraphics[width=.7\textwidth]{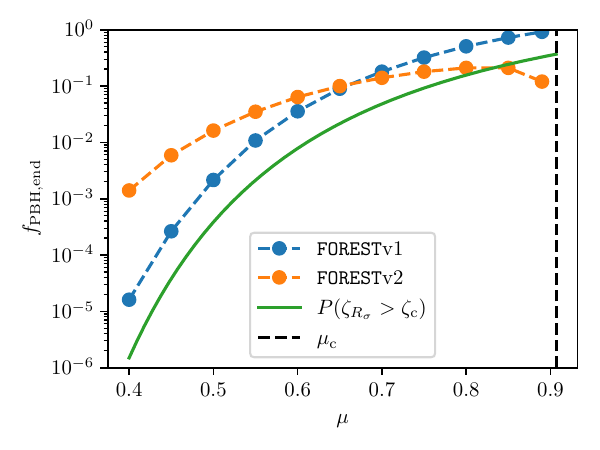}
    \caption{Comparison between the coarse-shelled proxy of ref.~\cite{Animali:2025pyf} (\texttt{FOREST}v1) and the compaction function approach developed in the present work (\texttt{FOREST}v2) on the mass fraction of PBHs, $f_\mathrm{PBH,end}$, for the flat well corresponding to $d=0$.}
    \label{fig:fraction}
\end{figure}

\begin{figure}
    \centering
    \begin{subfigure}{.45\textwidth}
        \includegraphics[width=\textwidth]{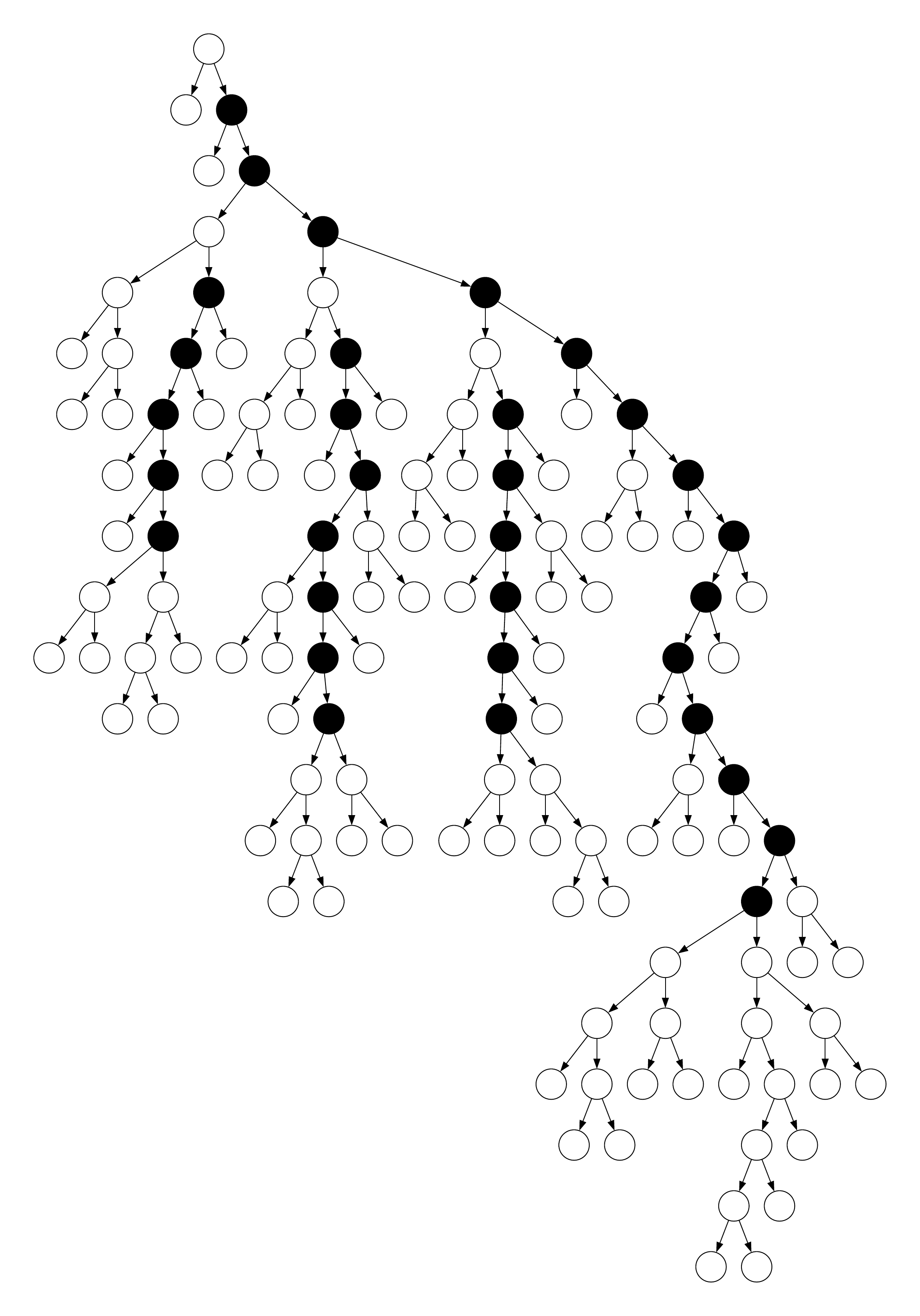}
        \caption{\texttt{FOREST}v1}
    \end{subfigure}
    \begin{subfigure}{.45\textwidth}
        \includegraphics[width=\textwidth]{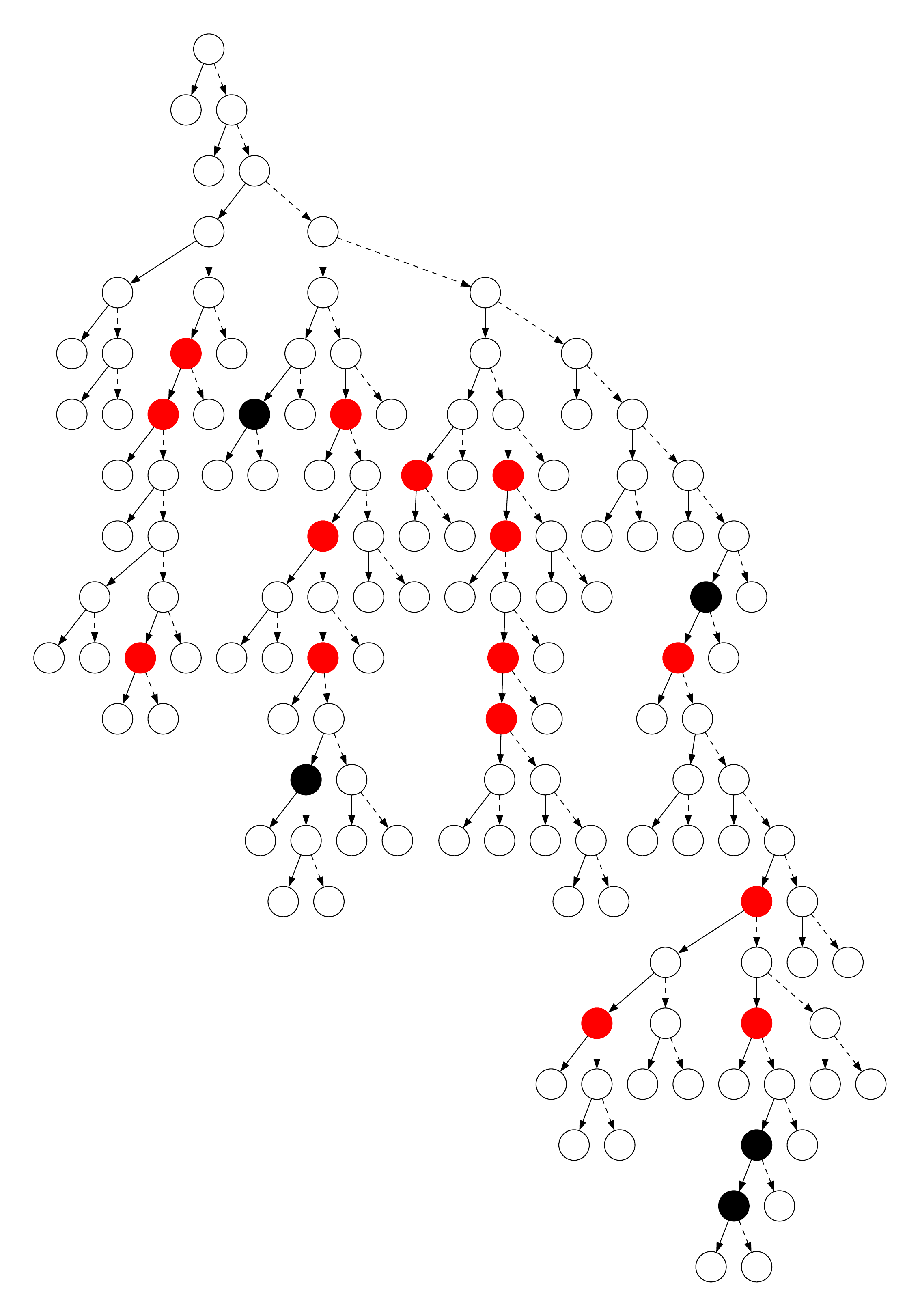}
        \caption{\texttt{FOREST}v2}
    \end{subfigure}
    \caption{Comparison between the two prescriptions for an example tree. For \texttt{FORESTv1}, all PBH-forming nodes are colored black, and there is no distinction between left-hand and right-hand nodes. For \texttt{FORESTv2}, the conventions follow those of \cref{fig:numerical_example_tree}.}
    \label{fig:tree_comparison}
\end{figure}

In ref.~\cite{Animali:2025pyf}, a criterion for primordial black hole formation was formulated in terms of the coarse-shelled curvature perturbation, introduced as a proxy for the compaction function. The underlying idea, which we briefly review now, is that the difference between the curvature perturbation coarse-grained at two distinct radii isolates fluctuations around the scale of interest while suppressing large-scale contributions, thereby providing a more local tracer of overdense regions. Within the stochastic-tree framework, this prescription amounts to computing the curvature perturbation at a given node relative to its local environment, represented by its immediate parent. Taking the node $b_\mathrm{L}$ in \cref{fig:nodes_in_derivatives} as a reference, we can define $\zeta_{b_\mathrm{L}\,a}=\zeta_{b_\mathrm{L}}-\zeta_a$. By matching the effective window functions of this coarse-shelled proxy and of the compaction function, ref.~\cite{Animali:2025pyf} mapped the collapse threshold on the latter onto an equivalent condition for the former. During radiation domination, adopting a collapse threshold $\mathcal{C_\mathrm{c}}=1/2$ for the compaction function --- as we also assume in the present work, see \cref{sec:pbh:threshold_I_II} --- this procedure leads to the condition $\zeta_{b_\mathrm{L}\,a,\mathrm{c}}=1/2 \ln{(V_a/V_{b_{\mathrm{L}}})}$. 
All nodes satisfying the condition $\zeta_{b_\mathrm{L}\,a}>\zeta_{b_\mathrm{L}\,a,\mathrm{c}}$ are then assumed to collapse into PBHs. The latter condition can also be recast as $\zeta_{b_\mathrm{L}\,a}/\zeta_{b_\mathrm{L}\,a,\mathrm{c}}\simeq 2 (W_{b_\mathrm{L}}-W_{b_\mathrm{R}})>1$, where $W$ denotes the volume-averaged number of \efolds:\footnote{In stochastic trees, the volume-averaged number of \efolds $W$ corresponds to the number of \efolds realized from a certain node, volume averaged over the set of child leaves emerging from it~\cite{Animali:2024jiz, Animali:2025pyf}.} when $V_{b_\mathrm{L}}\gg V_{b_\mathrm{R}}$, in most cases $W_{b_\mathrm{L}}\gg W_{b_\mathrm{R}}$ and the condition $\zeta_{b_{\mathrm{L}}\,a}>\zeta_{b_{\mathrm{L}}\,a,\mathrm{c}}$ is satisfied. This corresponds to highly asymmetric configurations in which two sibling nodes generate very different volumes.

It is worth comparing the predictions obtained from this old coarse-shelled proxy with those obtained from the new compaction-function prescription.
For a quantitative comparison, we focus on the predicted PBH abundance for $d=0$, corresponding to the flat-well scenario previously analysed in ref.~\cite{Animali:2025pyf} within the stochastic-tree framework. This case may be regarded as conservative. Indeed, since the flat well is the limiting case of a tilted well in the quantum regime, it probes both the quantum and the near-critical regimes as can be seen in \cref{fig:phase_diagram}. Hence, it also represents the most extreme situation in terms of large fluctuations, PBH formation and cloud-in-cloud, which are all features enhancing the upward propagation of PBH-forming nodes within stochastic trees. One may therefore expect the agreement between the two prescriptions to become only better when a non-zero tilt is included.

In \cref{fig:fraction}, we show $f_{\mathrm{PBH,end}}$, namely, the fraction of the universe at the end of inflation that eventually collapses into PBHs, as a function of $\mu$, for $d=0$. The blue points correspond to stochastic-tree simulations based on the coarse-shelled proxy, while the orange points are obtained using the compaction function criterion developed in the present work, denoted as \texttt{FOREST}v1 and \texttt{FOREST}v2 respectively. For comparison, the green curve represents the simple Press--Schechter estimate based on the curvature perturbation $P[\zeta>\zeta_{\mathrm{c}}]=\int_{\zeta_{\mathrm{c}}}^\infty P(\zeta) \mathrm{d}\zeta$ with $\zeta_{\mathrm{c}}=1$ and $P(\zeta)$ the one-point distribution of the curvature perturbation coarse-grained at the end of inflation in the flat well, for which the exact analytical solution is known~\cite{Animali:2024jiz}. As already noted in ref.~\cite{Animali:2025pyf}, the Press–Schechter estimate and the coarse-shelled-proxy result exhibit the same qualitative behaviour and agree quantitatively to within about one order of magnitude. By contrast, the compaction-function-based result is qualitatively different. For large values of $\mu$, the compaction function prediction lies below the coarse-shelled one. In particular, as $\mu$ approaches the critical value for eternal inflation, marked by the black dashed line, the coarse-shelled proxy yields $f_{\mathrm{PBH},\mathrm{end}}\rightarrow 1$. This suggests that the forest becomes dominated by a small number of very large branches that collapse into PBHs. By contrast, the compaction function prediction remains smaller. This is a desirable outcome and follows from the PBH criterion introduced in this work, which therefore seems to suppress the undesired tendency, observed with the coarse-shelled proxy, for the collapse condition to propagate upwards until the root in the case of large fluctuations.

As $\mu$ decreases, the behaviour reverses, and below a certain value the orange curve lies above the blue one. This regime should, however, be interpreted with caution: for small $\mu$ the stochastic trees contain fewer nodes and our framework becomes more sensitive to discretization artefacts,
potentially leading to an overestimation of the PBH abundance.

The comparison can also be made at the level of individual tree realizations, see \cref{fig:tree_comparison}.
One immediately notices the tendency of upward propagation in the left panel, where the tree contains a single gigantic PBH-forming region. This tendency is mitigated in the right panel, where the tree instead contains a small number of separated PBHs. Moreover, while the scheme on the left is blind to the type of fluctuations (and thus PBHs), the prescription shown on the right distinguishes between type-I (black) and type-II (red) PBH forming nodes.

\section{Conclusion}
\label{sec:conclusion}

In this work, we have shown how the compaction function can be computed at large scales and at the end of inflation in the stochastic-inflation formalism. This involves solving a branching process for the random-field evolution along stochastic binary trees, and recording the volume that emerges from each node. The recursive structure underlying stochastic trees can be readily implemented, and we provide the publicly-available numerical tool \texttt{FOREST}~\cite{auclair_2025_15235932}, introduced in ref.~\cite{Animali:2025pyf} and extended in the present work to include the computation of the compaction function. This automatically includes cloud-in-cloud effects (namely the fact that PBHs may form in regions that already contain smaller PBHs). The compaction function has become the state-of-the-art diagnosis for primordial black hole formation. This work thus connects the main non-perturbative approach to inflationary fluctuations, stochastic inflation, to refined methods for PBH mass fraction estimates, thereby advancing the ongoing effort to build a comprehensive pipeline that extracts PBH properties from high-energy models of the early universe. The main results obtained in this work are the following:

\begin{itemize}
\item By using an appropriate discretization procedure (namely one that is compatible with the symmetries underlying the compaction function and that recovers the main limits of interest, ensuring consistency with the continuum limit), we have shown how the compaction function can be extracted from the ratio of the volume emerging from the sibling and the right child of a given node in the binary tree, see \cref{eq:Cl_offbranch}. 
\item This diagnosis also allowed us to determine whether the areal radius of the perturbation increases monotonically with the radial coordinate or not, which separates so-called type-I from type-II fluctuations. Type-I fluctuations above the collapse threshold tend to form an apparent horizon when crossing back inside the Hubble radius, giving rise to ``standard'' primordial black holes. Type-II fluctuations may give rise to a ``separate universe'' made of an inflating bubble, connected to the Friedmann-Lema\^itre bulk through a throat. The fate of type-II fluctuations is however less clear, and this is why in this work we performed a separate census of type-I and type-II fluctuations on the end-of-inflation hypersurface, while remaining agnostic about their subsequent gravitational evolution.
\item For illustration, we have applied our framework to a single-field slow-roll model of inflation where the potential has a constant slope over a finite field interval, outside which quantum diffusion is neglected, and which is thus bounded by one reflective and one absorbing boundary condition. The advantage of this model is that it contains enough parameters to probe (i) the \emph{classical} limit, where quantum diffusion is negligible, fluctuations are quasi-Gaussian and PBH abundances are low; and (ii) the \emph{quantum} limit in which the potential drift is sub-dominant, fluctuations feature heavy-tailed statistics and PBHs are copiously produced.
\item We identified a region in parameter space where the mean volume diverges. In this ``eternal-inflation'' regime, our program cannot be applied since the volume statistics do not converge, so we restricted our analysis to outside this domain. We nonetheless probed the \emph{near-critical} regime that lies close to the eternal-inflation boundary (see \cref{fig:phase_diagram}).
\item In the classical regime, the PBH mass function is narrow and the overall abundance is suppressed, although cloud-in-cloud remains relevant. Type-II fluctuations are rare and
most PBHs proceed from type-I fluctuations, as usually argued.
\item In the quantum and near-critical regimes, the PBH mass function extends over several orders of magnitude and the total PBH abundance is substantial. The type-II population is less suppressed and they may even outnumber type-I PBHs. When this is the case, the type-II mass distribution is broader than the one of type I. 
\item We compared our results with the coarse-shelled proxy for the compaction function that we have previously proposed. While they are qualitatively similar, in the quantum and near-critical regimes the coarse-shelled scheme may lead to spurious upward propagation of black holes in the tree, since it is more sensitive to global imbalance. This leads to quantitative differences with the present approach, which is better suited to tackle these regimes. The coarse-shelled method is also unable to discriminate type-I from type-II fluctuations. 
\end{itemize}

The most important result is maybe the fact that, in regimes leading to abundant PBH formation, type-II fluctuations become as frequent as type-I fluctuations. This implies that, not only the gravitational properties of type-II fluctuations needs to be further understood as their abundance may be substantial, but their interplay with other engulfing or engulfed type-I/II fluctuations should also be studied. Indeed, the abundant-PBH regime also corresponds to where cloud-in-cloud effects are important. While the compaction function has mostly been applied to isolated peaks in the literature, further analysis of compaction functions with several overcritical maxima is thus required. Note also that, in the tree picture, a type-II perturbation may contain a significant volume of leaves, and it is not clear how this volume should contribute to compaction function computations at longer distances, higher up along the branch. Indeed, if a type-II perturbation collapses into a black hole, it could be argued that the volume it contains should stay hidden behind the event horizon and not be propagated upward, possibly reducing PBH formation at longer distances. We hope to return to such questions in future work.

This work also shows how the stochastic-tree approach can be enriched without losing its numerical efficiency and intuitive features. The formalism could be further expanded by extracting real-space maps from the stochastic trees. Endowing the trees with a consistent geometry is challenging --- in particular, the local picture made of concentric shells in \cref{sec:compact:trees} assumes spherical symmetry, which obviously cannot be extended to the whole spacetime with multiple competing branches. Still, local relations may provide useful insights into how the topological tree structure could be mapped to a full spacetime metric. We leave the details of such a mapping for future work. Even in the absence of a full map, the spatial correlations of local features can be analysed, at least when the features are rare enough so that the bulk of spacetime can be approximated as FLRW. In a forthcoming article, we will present quantitative results about PBH clustering in such a setup,  when strong non-Gaussian effects are present. From the knowledge of the compaction function, the statistics of the recently proposed primordial voids~\cite{Joana:2025gqf} could also be worked out.

Finally, it would be interesting to use the tools introduced in this work to study more realistic scenarios of inflation, in particular those that include a phase of ultra-slow roll~\cite{Pattison:2019hef, Firouzjahi:2018vet, Figueroa:2020jkf, Figueroa:2021zah, Mishra:2023lhe}. This would require keeping track of leading gradient effects during the slow-roll--ultra-slow-roll transition~\cite{Jackson:2023obv, Artigas:2024ajh, Raveendran:2025pnz, Briaud:2025ayt}.

\begin{acknowledgments}
    The authors thank Christophe Ringeval for insightful discussions.
	Baptiste Blachier is a Research Fellow of the Fonds de la Recherche Scientifique – FNRS.
	Chiara Animali is supported by the ESA Belgian Federal PRODEX
	Grant $\mathrm{N^{\circ}} 4000143201$.
	Eemeli Tomberg is supported by the ``Fonds de la Recherche Scientifique'' (FNRS) under the IISN grant number 4.4517.08.
	Computational resources have been provided by the CURL's development cluster, the supercomputing facilities of the Universit\'e catholique de Louvain (CISM/UCL) and the Consortium des \'Equipements de Calcul Intensif en F\'ed\'eration Wallonie Bruxelles (C\'ECI) funded by the Fond de la Recherche Scientifique de Belgique (F.R.S.-FNRS) under convention 2.5020.11 and by the Walloon Region.
\end{acknowledgments}

\bibliographystyle{JHEP}
\bibliography{TreeCompaction}

\providecommand{\href}[2]{#2}\begingroup\raggedright\begin{thebibliography}{10}

\bibitem{SDSS:2005xqv}
{\scshape SDSS} collaboration, D.~J. Eisenstein et~al., \emph{{Detection of the
  Baryon Acoustic Peak in the Large-Scale Correlation Function of SDSS Luminous
  Red Galaxies}}, \href{http://dx.doi.org/10.1086/466512}{\emph{Astrophys. J.}
  {\bf 633} (2005) 560--574},
  [\href{http://arxiv.org/abs/astro-ph/0501171}{{\tt astro-ph/0501171}}].

\bibitem{Planck:2018nkj}
{\scshape Planck} collaboration, N.~Aghanim et~al., \emph{{Planck 2018 results.
  I. Overview and the cosmological legacy of Planck}},
  \href{http://dx.doi.org/10.1051/0004-6361/201833880}{\emph{Astron.
  Astrophys.} {\bf 641} (2020) A1},
  [\href{http://arxiv.org/abs/1807.06205}{{\tt 1807.06205}}].

\bibitem{Ivanov:2019pdj}
M.~M. Ivanov, M.~Simonovi\'c and M.~Zaldarriaga, \emph{{Cosmological Parameters
  from the BOSS Galaxy Power Spectrum}},
  \href{http://dx.doi.org/10.1088/1475-7516/2020/05/042}{\emph{JCAP} {\bf 05}
  (2020) 042}, [\href{http://arxiv.org/abs/1909.05277}{{\tt 1909.05277}}].

\bibitem{Hawking:1971ei}
S.~Hawking, \emph{{Gravitationally collapsed objects of very low mass}},
  \href{http://dx.doi.org/10.1093/mnras/152.1.75}{\emph{Mon. Not. Roy. Astron.
  Soc.} {\bf 152} (1971) 75}.

\bibitem{Carr:1974nx}
B.~J. Carr and S.~W. Hawking, \emph{{Black holes in the early Universe}},
  \href{http://dx.doi.org/10.1093/mnras/168.2.399}{\emph{Mon. Not. Roy. Astron.
  Soc.} {\bf 168} (1974) 399--415}.

\bibitem{Carr:1975qj}
B.~J. Carr, \emph{{The Primordial black hole mass spectrum}},
  \href{http://dx.doi.org/10.1086/153853}{\emph{Astrophys. J.} {\bf 201} (1975)
  1--19}.

\bibitem{Pattison:2017mbe}
C.~Pattison, V.~Vennin, H.~Assadullahi and D.~Wands, \emph{{Quantum diffusion
  during inflation and primordial black holes}},
  \href{http://dx.doi.org/10.1088/1475-7516/2017/10/046}{\emph{JCAP} {\bf 10}
  (2017) 046}, [\href{http://arxiv.org/abs/1707.00537}{{\tt 1707.00537}}].

\bibitem{Ezquiaga:2019ftu}
J.~M. Ezquiaga, J.~Garc\'\i{}a-Bellido and V.~Vennin, \emph{{The exponential
  tail of inflationary fluctuations: consequences for primordial black holes}},
  \href{http://dx.doi.org/10.1088/1475-7516/2020/03/029}{\emph{JCAP} {\bf 03}
  (2020) 029}, [\href{http://arxiv.org/abs/1912.05399}{{\tt 1912.05399}}].

\bibitem{Starobinsky:1986fx}
A.~A. Starobinsky, \emph{{Stochastic De Sitter (inflationary) stage in the
  early universe}},
  \href{http://dx.doi.org/10.1007/3-540-16452-9_6}{\emph{Lect. Notes Phys.}
  {\bf 246} (1986) 107--126}.

\bibitem{Salopek:1990jq}
D.~S. Salopek and J.~R. Bond, \emph{{Nonlinear evolution of long wavelength
  metric fluctuations in inflationary models}},
  \href{http://dx.doi.org/10.1103/PhysRevD.42.3936}{\emph{Phys. Rev. D} {\bf
  42} (1990) 3936--3962}.

\bibitem{Sasaki:1995aw}
M.~Sasaki and E.~D. Stewart, \emph{{A General analytic formula for the spectral
  index of the density perturbations produced during inflation}},
  \href{http://dx.doi.org/10.1143/PTP.95.71}{\emph{Prog. Theor. Phys.} {\bf 95}
  (1996) 71--78}, [\href{http://arxiv.org/abs/astro-ph/9507001}{{\tt
  astro-ph/9507001}}].

\bibitem{Wands:2000dp}
D.~Wands, K.~A. Malik, D.~H. Lyth and A.~R. Liddle, \emph{{A New approach to
  the evolution of cosmological perturbations on large scales}},
  \href{http://dx.doi.org/10.1103/PhysRevD.62.043527}{\emph{Phys. Rev. D} {\bf
  62} (2000) 043527}, [\href{http://arxiv.org/abs/astro-ph/0003278}{{\tt
  astro-ph/0003278}}].

\bibitem{Lyth:2003im}
D.~H. Lyth and D.~Wands, \emph{{Conserved cosmological perturbations}},
  \href{http://dx.doi.org/10.1103/PhysRevD.68.103515}{\emph{Phys. Rev. D} {\bf
  68} (2003) 103515}, [\href{http://arxiv.org/abs/astro-ph/0306498}{{\tt
  astro-ph/0306498}}].

\bibitem{Rigopoulos:2003ak}
G.~I. Rigopoulos and E.~P.~S. Shellard, \emph{{The separate universe approach
  and the evolution of nonlinear superhorizon cosmological perturbations}},
  \href{http://dx.doi.org/10.1103/PhysRevD.68.123518}{\emph{Phys. Rev. D} {\bf
  68} (2003) 123518}, [\href{http://arxiv.org/abs/astro-ph/0306620}{{\tt
  astro-ph/0306620}}].

\bibitem{Lyth:2005fi}
D.~H. Lyth and Y.~Rodriguez, \emph{{The Inflationary prediction for primordial
  non-Gaussianity}},
  \href{http://dx.doi.org/10.1103/PhysRevLett.95.121302}{\emph{Phys. Rev.
  Lett.} {\bf 95} (2005) 121302},
  [\href{http://arxiv.org/abs/astro-ph/0504045}{{\tt astro-ph/0504045}}].

\bibitem{Artigas:2021zdk}
D.~Artigas, J.~Grain and V.~Vennin, \emph{{Hamiltonian formalism for
  cosmological perturbations: the~separate-universe approach}},
  \href{http://dx.doi.org/10.1088/1475-7516/2022/02/001}{\emph{JCAP} {\bf 02}
  (2022) 001}, [\href{http://arxiv.org/abs/2110.11720}{{\tt 2110.11720}}].

\bibitem{Jackson:2023obv}
J.~H.~P. Jackson, H.~Assadullahi, A.~D. Gow, K.~Koyama, V.~Vennin and D.~Wands,
  \emph{{The separate-universe approach and sudden transitions during
  inflation}},
  \href{http://dx.doi.org/10.1088/1475-7516/2024/05/053}{\emph{JCAP} {\bf 05}
  (2024) 053}, [\href{http://arxiv.org/abs/2311.03281}{{\tt 2311.03281}}].

\bibitem{Vennin:2015hra}
V.~Vennin and A.~A. Starobinsky, \emph{{Correlation Functions in Stochastic
  Inflation}},
  \href{http://dx.doi.org/10.1140/epjc/s10052-015-3643-y}{\emph{Eur. Phys. J.
  C} {\bf 75} (2015) 413}, [\href{http://arxiv.org/abs/1506.04732}{{\tt
  1506.04732}}].

\bibitem{Lesgourgues:1996jc}
J.~Lesgourgues, D.~Polarski and A.~A. Starobinsky, \emph{{Quantum to classical
  transition of cosmological perturbations for nonvacuum initial states}},
  \href{http://dx.doi.org/10.1016/S0550-3213(97)00224-1}{\emph{Nucl. Phys. B}
  {\bf 497} (1997) 479--510}, [\href{http://arxiv.org/abs/gr-qc/9611019}{{\tt
  gr-qc/9611019}}].

\bibitem{Grain:2017dqa}
J.~Grain and V.~Vennin, \emph{{Stochastic inflation in phase space: Is slow
  roll a stochastic attractor?}},
  \href{http://dx.doi.org/10.1088/1475-7516/2017/05/045}{\emph{JCAP} {\bf 05}
  (2017) 045}, [\href{http://arxiv.org/abs/1703.00447}{{\tt 1703.00447}}].

\bibitem{Starobinsky:1982ee}
A.~A. Starobinsky, \emph{{Dynamics of Phase Transition in the New Inflationary
  Universe Scenario and Generation of Perturbations}},
  \href{http://dx.doi.org/10.1016/0370-2693(82)90541-X}{\emph{Phys. Lett. B}
  {\bf 117} (1982) 175--178}.

\bibitem{Starobinsky:1985ibc}
A.~A. Starobinsky, \emph{{Multicomponent de Sitter (Inflationary) Stages and
  the Generation of Perturbations}}, {\emph{JETP Lett.} {\bf 42} (1985)
  152--155}.

\bibitem{Sasaki:1998ug}
M.~Sasaki and T.~Tanaka, \emph{{Superhorizon scale dynamics of multiscalar
  inflation}}, \href{http://dx.doi.org/10.1143/PTP.99.763}{\emph{Prog. Theor.
  Phys.} {\bf 99} (1998) 763--782},
  [\href{http://arxiv.org/abs/gr-qc/9801017}{{\tt gr-qc/9801017}}].

\bibitem{Lyth:2004gb}
D.~H. Lyth, K.~A. Malik and M.~Sasaki, \emph{{A General proof of the
  conservation of the curvature perturbation}},
  \href{http://dx.doi.org/10.1088/1475-7516/2005/05/004}{\emph{JCAP} {\bf 05}
  (2005) 004}, [\href{http://arxiv.org/abs/astro-ph/0411220}{{\tt
  astro-ph/0411220}}].

\bibitem{Finelli:2008zg}
F.~Finelli, G.~Marozzi, A.~Starobinsky, G.~Vacca and G.~Venturi,
  \emph{{Generation of fluctuations during inflation: Comparison of stochastic
  and field-theoretic approaches}},
  \href{http://dx.doi.org/10.1103/PhysRevD.79.044007}{\emph{Phys. Rev. D} {\bf
  79} (2009) 044007}, [\href{http://arxiv.org/abs/0808.1786}{{\tt 0808.1786}}].

\bibitem{Finelli:2010sh}
F.~Finelli, G.~Marozzi, A.~A. Starobinsky, G.~P. Vacca and G.~Venturi,
  \emph{{Stochastic growth of quantum fluctuations during slow-roll
  inflation}}, \href{http://dx.doi.org/10.1103/PhysRevD.82.064020}{\emph{Phys.
  Rev. D} {\bf 82} (2010) 064020}, [\href{http://arxiv.org/abs/1003.1327}{{\tt
  1003.1327}}].

\bibitem{Enqvist:2008kt}
K.~Enqvist, S.~Nurmi, D.~Podolsky and G.~Rigopoulos, \emph{{On the divergences
  of inflationary superhorizon perturbations}},
  \href{http://dx.doi.org/10.1088/1475-7516/2008/04/025}{\emph{JCAP} {\bf 04}
  (2008) 025}, [\href{http://arxiv.org/abs/0802.0395}{{\tt 0802.0395}}].

\bibitem{Fujita:2013cna}
T.~Fujita, M.~Kawasaki, Y.~Tada and T.~Takesako, \emph{{A new algorithm for
  calculating the curvature perturbations in stochastic inflation}},
  \href{http://dx.doi.org/10.1088/1475-7516/2013/12/036}{\emph{JCAP} {\bf 12}
  (2013) 036}, [\href{http://arxiv.org/abs/1308.4754}{{\tt 1308.4754}}].

\bibitem{Panagopoulos:2019ail}
G.~Panagopoulos and E.~Silverstein, \emph{{Primordial Black Holes from
  non-Gaussian tails}},  \href{http://arxiv.org/abs/1906.02827}{{\tt
  1906.02827}}.

\bibitem{Figueroa:2020jkf}
D.~G. Figueroa, S.~Raatikainen, S.~Rasanen and E.~Tomberg, \emph{{Non-Gaussian
  Tail of the Curvature Perturbation in Stochastic Ultraslow-Roll Inflation:
  Implications for Primordial Black Hole Production}},
  \href{http://dx.doi.org/10.1103/PhysRevLett.127.101302}{\emph{Phys. Rev.
  Lett.} {\bf 127} (2021) 101302}, [\href{http://arxiv.org/abs/2012.06551}{{\tt
  2012.06551}}].

\bibitem{Figueroa:2021zah}
D.~G. Figueroa, S.~Raatikainen, S.~Rasanen and E.~Tomberg, \emph{{Implications
  of stochastic effects for primordial black hole production in ultra-slow-roll
  inflation}},
  \href{http://dx.doi.org/10.1088/1475-7516/2022/05/027}{\emph{JCAP} {\bf 05}
  (2022) 027}, [\href{http://arxiv.org/abs/2111.07437}{{\tt 2111.07437}}].

\bibitem{Pattison:2021oen}
C.~Pattison, V.~Vennin, D.~Wands and H.~Assadullahi, \emph{{Ultra-slow-roll
  inflation with quantum diffusion}},
  \href{http://dx.doi.org/10.1088/1475-7516/2021/04/080}{\emph{JCAP} {\bf 04}
  (2021) 080}, [\href{http://arxiv.org/abs/2101.05741}{{\tt 2101.05741}}].

\bibitem{Tomberg:2021xxv}
E.~Tomberg, \emph{{A numerical approach to stochastic inflation and primordial
  black holes}},
  \href{http://dx.doi.org/10.1088/1742-6596/2156/1/012010}{\emph{J. Phys. Conf.
  Ser.} {\bf 2156} (2021) 012010}, [\href{http://arxiv.org/abs/2110.10684}{{\tt
  2110.10684}}].

\bibitem{Rigopoulos:2021nhv}
G.~Rigopoulos and A.~Wilkins, \emph{{Inflation is always semi-classical:
  diffusion domination overproduces Primordial Black Holes}},
  \href{http://dx.doi.org/10.1088/1475-7516/2021/12/027}{\emph{JCAP} {\bf 12}
  (2021) 027}, [\href{http://arxiv.org/abs/2107.05317}{{\tt 2107.05317}}].

\bibitem{Achucarro:2021pdh}
A.~Achucarro, S.~Cespedes, A.-C. Davis and G.~A. Palma, \emph{{The hand-made
  tail: non-perturbative tails from multifield inflation}},
  \href{http://dx.doi.org/10.1007/JHEP05(2022)052}{\emph{JHEP} {\bf 05} (2022)
  052}, [\href{http://arxiv.org/abs/2112.14712}{{\tt 2112.14712}}].

\bibitem{Ezquiaga:2022qpw}
J.~M. Ezquiaga, J.~Garc\'\i{}a-Bellido and V.~Vennin, \emph{{Massive Galaxy
  Clusters Like El Gordo Hint at Primordial Quantum Diffusion}},
  \href{http://dx.doi.org/10.1103/PhysRevLett.130.121003}{\emph{Phys. Rev.
  Lett.} {\bf 130} (2023) 121003}, [\href{http://arxiv.org/abs/2207.06317}{{\tt
  2207.06317}}].

\bibitem{Animali:2022otk}
C.~Animali and V.~Vennin, \emph{{Primordial black holes from stochastic
  tunnelling}},
  \href{http://dx.doi.org/10.1088/1475-7516/2023/02/043}{\emph{JCAP} {\bf 02}
  (2023) 043}, [\href{http://arxiv.org/abs/2210.03812}{{\tt 2210.03812}}].

\bibitem{Cai:2022erk}
Y.-F. Cai, X.-H. Ma, M.~Sasaki, D.-G. Wang and Z.~Zhou, \emph{{Highly
  non-Gaussian tails and primordial black holes from single-field inflation}},
  \href{http://dx.doi.org/10.1088/1475-7516/2022/12/034}{\emph{JCAP} {\bf 12}
  (2022) 034}, [\href{http://arxiv.org/abs/2207.11910}{{\tt 2207.11910}}].

\bibitem{Gow:2022jfb}
A.~D. Gow, H.~Assadullahi, J.~H.~P. Jackson, K.~Koyama, V.~Vennin and D.~Wands,
  \emph{{Non-perturbative non-Gaussianity and primordial black holes}},
  \href{http://dx.doi.org/10.1209/0295-5075/acd417}{\emph{EPL} {\bf 142} (2023)
  49001}, [\href{http://arxiv.org/abs/2211.08348}{{\tt 2211.08348}}].

\bibitem{Tomberg:2023kli}
E.~Tomberg, \emph{{Stochastic constant-roll inflation and primordial black
  holes}}, \href{http://dx.doi.org/10.1103/PhysRevD.108.043502}{\emph{Phys.
  Rev. D} {\bf 108} (2023) 043502},
  [\href{http://arxiv.org/abs/2304.10903}{{\tt 2304.10903}}].

\bibitem{Briaud:2023eae}
V.~Briaud and V.~Vennin, \emph{{Uphill inflation}},
  \href{http://dx.doi.org/10.1088/1475-7516/2023/06/029}{\emph{JCAP} {\bf 06}
  (2023) 029}, [\href{http://arxiv.org/abs/2301.09336}{{\tt 2301.09336}}].

\bibitem{Vennin:2024yzl}
V.~Vennin and D.~Wands, \emph{{Quantum Diffusion and~Large Primordial
  Perturbations from~Inflation}}.
\newblock 2025.
\newblock \href{http://arxiv.org/abs/2402.12672}{{\tt 2402.12672}}.
\newblock 10.1007/978-981-97-8887-3\_8.

\bibitem{Inui:2024sce}
R.~Inui, H.~Motohashi, S.~Pi, Y.~Tada and S.~Yokoyama, \emph{{Constant roll and
  non-Gaussian tail in light of logarithmic duality}},
  \href{http://dx.doi.org/10.1088/1475-7516/2025/02/042}{\emph{JCAP} {\bf 02}
  (2025) 042}, [\href{http://arxiv.org/abs/2409.13500}{{\tt 2409.13500}}].

\bibitem{Sharma:2024fbr}
D.~Sharma, \emph{{Stochastic inflation and non-perturbative power spectrum
  beyond slow roll}},
  \href{http://dx.doi.org/10.1088/1475-7516/2025/03/017}{\emph{JCAP} {\bf 03}
  (2025) 017}, [\href{http://arxiv.org/abs/2411.08854}{{\tt 2411.08854}}].

\bibitem{Ando:2020fjm}
K.~Ando and V.~Vennin, \emph{{Power spectrum in stochastic inflation}},
  \href{http://dx.doi.org/10.1088/1475-7516/2021/04/057}{\emph{JCAP} {\bf 04}
  (2021) 057}, [\href{http://arxiv.org/abs/2012.02031}{{\tt 2012.02031}}].

\bibitem{Tada:2021zzj}
Y.~Tada and V.~Vennin, \emph{{Statistics of coarse-grained cosmological fields
  in stochastic inflation}},
  \href{http://dx.doi.org/10.1088/1475-7516/2022/02/021}{\emph{JCAP} {\bf 02}
  (2022) 021}, [\href{http://arxiv.org/abs/2111.15280}{{\tt 2111.15280}}].

\bibitem{Animali:2024jiz}
C.~Animali and V.~Vennin, \emph{{Clustering of primordial black holes from
  quantum diffusion during inflation}},
  \href{http://dx.doi.org/10.1088/1475-7516/2024/08/026}{\emph{JCAP} {\bf 08}
  (2024) 026}, [\href{http://arxiv.org/abs/2402.08642}{{\tt 2402.08642}}].

\bibitem{Animali:2025pyf}
C.~Animali, P.~Auclair, B.~Blachier and V.~Vennin, \emph{{Harvesting primordial
  black holes from stochastic trees with FOREST}},
  \href{http://dx.doi.org/10.1088/1475-7516/2025/05/019}{\emph{JCAP} {\bf 05}
  (2025) 019}, [\href{http://arxiv.org/abs/2501.05371}{{\tt 2501.05371}}].

\bibitem{Linde:1993xx}
A.~D. Linde, D.~A. Linde and A.~Mezhlumian, \emph{{From the Big Bang theory to
  the theory of a stationary universe}},
  \href{http://dx.doi.org/10.1103/PhysRevD.49.1783}{\emph{Phys. Rev. D} {\bf
  49} (1994) 1783--1826}, [\href{http://arxiv.org/abs/gr-qc/9306035}{{\tt
  gr-qc/9306035}}].

\bibitem{Jain:2019gsq}
M.~Jain and M.~P. Hertzberg, \emph{{Statistics of Inflating Regions in Eternal
  Inflation}}, \href{http://dx.doi.org/10.1103/PhysRevD.100.023513}{\emph{Phys.
  Rev. D} {\bf 100} (2019) 023513},
  [\href{http://arxiv.org/abs/1904.04262}{{\tt 1904.04262}}].

\bibitem{auclair_2025_15235932}
P.~Auclair, \emph{Forest: Fortran recursive exploration of stochastic trees},
  Apr., 2025.
\newblock 10.5281/zenodo.15235932.

\bibitem{Jedamzik:1994nr}
K.~Jedamzik, \emph{{The Cloud in cloud problem in the Press-Schechter formalism
  of hierarchical structure formation}},
  \href{http://dx.doi.org/10.1086/175936}{\emph{Astrophys. J.} {\bf 448} (1995)
  1--17}, [\href{http://arxiv.org/abs/astro-ph/9408080}{{\tt
  astro-ph/9408080}}].

\bibitem{Shibata:1999zs}
M.~Shibata and M.~Sasaki, \emph{{Black hole formation in the Friedmann
  universe: Formulation and computation in numerical relativity}},
  \href{http://dx.doi.org/10.1103/PhysRevD.60.084002}{\emph{Phys. Rev. D} {\bf
  60} (1999) 084002}, [\href{http://arxiv.org/abs/gr-qc/9905064}{{\tt
  gr-qc/9905064}}].

\bibitem{Niemeyer:1997mt}
J.~C. Niemeyer and K.~Jedamzik, \emph{{Near-critical gravitational collapse and
  the initial mass function of primordial black holes}},
  \href{http://dx.doi.org/10.1103/PhysRevLett.80.5481}{\emph{Phys. Rev. Lett.}
  {\bf 80} (1998) 5481--5484},
  [\href{http://arxiv.org/abs/astro-ph/9709072}{{\tt astro-ph/9709072}}].

\bibitem{Musco:2004ak}
I.~Musco, J.~C. Miller and L.~Rezzolla, \emph{{Computations of primordial black
  hole formation}},
  \href{http://dx.doi.org/10.1088/0264-9381/22/7/013}{\emph{Class. Quant.
  Grav.} {\bf 22} (2005) 1405--1424},
  [\href{http://arxiv.org/abs/gr-qc/0412063}{{\tt gr-qc/0412063}}].

\bibitem{Nakama:2013ica}
T.~Nakama, T.~Harada, A.~G. Polnarev and J.~Yokoyama, \emph{{Identifying the
  most crucial parameters of the initial curvature profile for primordial black
  hole formation}},
  \href{http://dx.doi.org/10.1088/1475-7516/2014/01/037}{\emph{JCAP} {\bf 01}
  (2014) 037}, [\href{http://arxiv.org/abs/1310.3007}{{\tt 1310.3007}}].

\bibitem{Harada:2013epa}
T.~Harada, C.-M. Yoo and K.~Kohri, \emph{{Threshold of primordial black hole
  formation}}, \href{http://dx.doi.org/10.1103/PhysRevD.88.084051}{\emph{Phys.
  Rev. D} {\bf 88} (2013) 084051}, [\href{http://arxiv.org/abs/1309.4201}{{\tt
  1309.4201}}].

\bibitem{Harada:2015yda}
T.~Harada, C.-M. Yoo, T.~Nakama and Y.~Koga, \emph{{Cosmological
  long-wavelength solutions and primordial black hole formation}},
  \href{http://dx.doi.org/10.1103/PhysRevD.91.084057}{\emph{Phys. Rev. D} {\bf
  91} (2015) 084057}, [\href{http://arxiv.org/abs/1503.03934}{{\tt
  1503.03934}}].

\bibitem{Musco:2018rwt}
I.~Musco, \emph{{Threshold for primordial black holes: Dependence on the shape
  of the cosmological perturbations}},
  \href{http://dx.doi.org/10.1103/PhysRevD.100.123524}{\emph{Phys. Rev. D} {\bf
  100} (2019) 123524}, [\href{http://arxiv.org/abs/1809.02127}{{\tt
  1809.02127}}].

\bibitem{Escriva:2019phb}
A.~Escriv\`a, C.~Germani and R.~K. Sheth, \emph{{Universal threshold for
  primordial black hole formation}},
  \href{http://dx.doi.org/10.1103/PhysRevD.101.044022}{\emph{Phys. Rev. D} {\bf
  101} (2020) 044022}, [\href{http://arxiv.org/abs/1907.13311}{{\tt
  1907.13311}}].

\bibitem{Musco:2020jjb}
I.~Musco, V.~De~Luca, G.~Franciolini and A.~Riotto, \emph{{Threshold for
  primordial black holes. II. A simple analytic prescription}},
  \href{http://dx.doi.org/10.1103/PhysRevD.103.063538}{\emph{Phys. Rev. D} {\bf
  103} (2021) 063538}, [\href{http://arxiv.org/abs/2011.03014}{{\tt
  2011.03014}}].

\bibitem{Escriva:2020tak}
A.~Escriv{\`a}, C.~Germani and R.~K. Sheth, \emph{{Analytical thresholds for
  black hole formation in general cosmological backgrounds}},
  \href{http://dx.doi.org/10.1088/1475-7516/2021/01/030}{\emph{JCAP} {\bf 01}
  (2021) 030}, [\href{http://arxiv.org/abs/2007.05564}{{\tt 2007.05564}}].

\bibitem{Raatikainen:2023bzk}
S.~Raatikainen, S.~R\"as\"anen and E.~Tomberg, \emph{{Primordial Black Hole
  Compaction Function from Stochastic Fluctuations in Ultraslow-Roll
  Inflation}},
  \href{http://dx.doi.org/10.1103/PhysRevLett.133.121403}{\emph{Phys. Rev.
  Lett.} {\bf 133} (2024) 121403}, [\href{http://arxiv.org/abs/2312.12911}{{\tt
  2312.12911}}].

\bibitem{Raatikainen:2025gpd}
S.~Raatikainen, S.~Rasanen and E.~Tomberg, \emph{{Effect of stochastic kicks on
  primordial black hole abundance and mass via the compaction function}},
  \href{http://dx.doi.org/10.1088/1475-7516/2026/03/063}{\emph{JCAP} {\bf 03}
  (2026) 063}, [\href{http://arxiv.org/abs/2510.09303}{{\tt 2510.09303}}].

\bibitem{Kopp:2010sh}
M.~Kopp, S.~Hofmann and J.~Weller, \emph{{Separate Universes Do Not Constrain
  Primordial Black Hole Formation}},
  \href{http://dx.doi.org/10.1103/PhysRevD.83.124025}{\emph{Phys. Rev. D} {\bf
  83} (2011) 124025}, [\href{http://arxiv.org/abs/1012.4369}{{\tt 1012.4369}}].

\bibitem{Carr:2014pga}
B.~J. Carr and T.~Harada, \emph{{Separate universe problem: 40 years on}},
  \href{http://dx.doi.org/10.1103/PhysRevD.91.084048}{\emph{Phys. Rev. D} {\bf
  91} (2015) 084048}, [\href{http://arxiv.org/abs/1405.3624}{{\tt 1405.3624}}].

\bibitem{Escriva:2023uko}
A.~Escriv{\`a}, V.~Atal and J.~Garriga, \emph{{Formation of trapped vacuum
  bubbles during inflation, and consequences for PBH scenarios}},
  \href{http://dx.doi.org/10.1088/1475-7516/2023/10/035}{\emph{JCAP} {\bf 10}
  (2023) 035}, [\href{http://arxiv.org/abs/2306.09990}{{\tt 2306.09990}}].

\bibitem{Harada:2024jxl}
T.~Harada, \emph{{Primordial Black Holes: Formation, Spin and Type II}},
  \href{http://dx.doi.org/10.3390/universe10120444}{\emph{Universe} {\bf 10}
  (2024) 444}, [\href{http://arxiv.org/abs/2409.01934}{{\tt 2409.01934}}].

\bibitem{Uehara:2024yyp}
K.~Uehara, A.~Escriv{\`a}, T.~Harada, D.~Saito and C.-M. Yoo, \emph{{Numerical
  simulation of type II primordial black hole formation}},
  \href{http://dx.doi.org/10.1088/1475-7516/2025/01/003}{\emph{JCAP} {\bf 01}
  (2025) 003}, [\href{http://arxiv.org/abs/2401.06329}{{\tt 2401.06329}}].

\bibitem{Escriva:2025eqc}
A.~Escriv{\`a}, \emph{{A new approach for simulating PBH formation from generic
  curvature fluctuations with the Misner-Sharp formalism}},
  \href{http://dx.doi.org/10.1016/j.dark.2025.102177}{\emph{Phys. Dark Univ.}
  {\bf 50} (2025) 102177}, [\href{http://arxiv.org/abs/2504.05813}{{\tt
  2504.05813}}].

\bibitem{Escriva:2025rja}
A.~Escriv{\`a}, \emph{{Threshold for PBH formation in the type-II region and
  its analytical estimation}},
  \href{http://dx.doi.org/10.1103/mq67-bbvj}{\emph{Phys. Rev. D} {\bf 112}
  (2025) 103527}, [\href{http://arxiv.org/abs/2504.05814}{{\tt 2504.05814}}].

\bibitem{Inui:2024fgk}
R.~Inui, C.~Joana, H.~Motohashi, S.~Pi, Y.~Tada and S.~Yokoyama,
  \emph{{Primordial black holes and induced gravitational waves from
  logarithmic non-Gaussianity}},
  \href{http://dx.doi.org/10.1088/1475-7516/2025/03/021}{\emph{JCAP} {\bf 03}
  (2025) 021}, [\href{http://arxiv.org/abs/2411.07647}{{\tt 2411.07647}}].

\bibitem{Uehara:2025idq}
K.~Uehara, A.~Escriv{\`a}, T.~Harada, D.~Saito and C.-M. Yoo, \emph{{Primordial
  black hole formation from a type II perturbation in the absence and presence
  of pressure}},
  \href{http://dx.doi.org/10.1088/1475-7516/2025/08/042}{\emph{JCAP} {\bf 08}
  (2025) 042}, [\href{http://arxiv.org/abs/2505.00366}{{\tt 2505.00366}}].

\bibitem{Escriva:2025ftp}
A.~Escriv{\`a}, J.~Garriga and S.~Pi, \emph{{Inflationary relics from an
  ultra-slow-roll plateau}},
  \href{http://dx.doi.org/10.1088/1475-7516/2026/03/018}{\emph{JCAP} {\bf 03}
  (2026) 018}, [\href{http://arxiv.org/abs/2512.04986}{{\tt 2512.04986}}].

\bibitem{Bardeen:1986}
J.~M. {Bardeen}, J.~R. {Bond}, N.~{Kaiser} and A.~S. {Szalay}, \emph{{The
  Statistics of Peaks of Gaussian Random Fields}},
  \href{http://dx.doi.org/10.1086/164143}{\emph{ApJ} {\bf 304} (1986) 15}.

\bibitem{Escriva:2024aeo}
A.~Escriv{\`a} and C.-M. Yoo, \emph{{Nonspherical effects on the mass function
  of primordial black holes}},
  \href{http://dx.doi.org/10.1103/4jbp-87wc}{\emph{Phys. Rev. D} {\bf 112}
  (2025) L081304}, [\href{http://arxiv.org/abs/2410.03451}{{\tt 2410.03451}}].

\bibitem{Escriva:2024lmm}
A.~Escriv{\`a} and C.-M. Yoo, \emph{{Simulations of ellipsoidal primordial
  black hole formation}},
  \href{http://dx.doi.org/10.1103/PhysRevD.112.083518}{\emph{Phys. Rev. D} {\bf
  112} (2025) 083518}, [\href{http://arxiv.org/abs/2410.03452}{{\tt
  2410.03452}}].

\bibitem{Germani:2023ojx}
C.~Germani and R.~K. Sheth, \emph{{The Statistics of Primordial Black Holes in
  a Radiation-Dominated Universe: Recent and New Results}},
  \href{http://dx.doi.org/10.3390/universe9090421}{\emph{Universe} {\bf 9}
  (2023) 421}, [\href{http://arxiv.org/abs/2308.02971}{{\tt 2308.02971}}].

\bibitem{Germani:2019zez}
C.~Germani and R.~K. Sheth, \emph{{Nonlinear statistics of primordial black
  holes from Gaussian curvature perturbations}},
  \href{http://dx.doi.org/10.1103/PhysRevD.101.063520}{\emph{Phys. Rev. D} {\bf
  101} (2020) 063520}, [\href{http://arxiv.org/abs/1912.07072}{{\tt
  1912.07072}}].

\bibitem{Harada:2024trx}
T.~Harada, H.~Iizuka, Y.~Koga and C.-M. Yoo, \emph{{Geometrical origin for the
  compaction function for primordial black hole formation}},
  \href{http://dx.doi.org/10.1103/PhysRevD.111.023537}{\emph{Phys. Rev. D} {\bf
  111} (2025) 023537}, [\href{http://arxiv.org/abs/2409.05544}{{\tt
  2409.05544}}].

\bibitem{Germani:2025hcu}
C.~Germani and L.~Montell{\`a}, \emph{{Trichotomy of primordial black holes
  initial conditions}}, \href{http://dx.doi.org/10.1103/6ysb-nbt8}{\emph{Phys.
  Rev. D} {\bf 113} (2026) 064054},
  [\href{http://arxiv.org/abs/2510.02006}{{\tt 2510.02006}}].

\bibitem{Shimada:2024eec}
M.~Shimada, A.~Escriv{\'a}, D.~Saito, K.~Uehara and C.-M. Yoo,
  \emph{{Primordial black hole formation from type II fluctuations with
  primordial non-Gaussianity}},
  \href{http://dx.doi.org/10.1088/1475-7516/2025/02/018}{\emph{JCAP} {\bf 02}
  (2025) 018}, [\href{http://arxiv.org/abs/2411.07648}{{\tt 2411.07648}}].

\bibitem{Fumagalli:2024kgg}
J.~Fumagalli, J.~Garriga, C.~Germani and R.~K. Sheth, \emph{{Unexpected shape
  of the primordial black hole mass function}},
  \href{http://dx.doi.org/10.1103/k75n-3qz4}{\emph{Phys. Rev. D} {\bf 111}
  (2025) 123518}, [\href{http://arxiv.org/abs/2412.07709}{{\tt 2412.07709}}].

\bibitem{Germani:2018jgr}
C.~Germani and I.~Musco, \emph{{Abundance of Primordial Black Holes Depends on
  the Shape of the Inflationary Power Spectrum}},
  \href{http://dx.doi.org/10.1103/PhysRevLett.122.141302}{\emph{Phys. Rev.
  Lett.} {\bf 122} (2019) 141302}, [\href{http://arxiv.org/abs/1805.04087}{{\tt
  1805.04087}}].

\bibitem{DeLuca:2020ioi}
V.~De~Luca, G.~Franciolini and A.~Riotto, \emph{{On the primordial black hole
  mass function for broad spectra}},
  \href{http://dx.doi.org/10.1016/j.physletb.2020.135550}{\emph{Phys. Lett. B}
  {\bf 807} (2020) 135550}, [\href{http://arxiv.org/abs/2001.04371}{{\tt
  2001.04371}}].

\bibitem{Auclair:2026kwf}
P.~Auclair, B.~Blachier and V.~Vennin, \emph{{Excursion set for primordial
  black holes: White noise and moving barriers}},
  \href{http://dx.doi.org/10.1103/clgf-sqv8}{\emph{Phys. Rev. D} {\bf 114}
  (2026) 063504}, [\href{http://arxiv.org/abs/2603.04185}{{\tt 2603.04185}}].

\bibitem{Nakama:2014fra}
T.~Nakama, \emph{{The double formation of primordial black holes}},
  \href{http://dx.doi.org/10.1088/1475-7516/2014/10/040}{\emph{JCAP} {\bf 10}
  (2014) 040}, [\href{http://arxiv.org/abs/1408.0955}{{\tt 1408.0955}}].

\bibitem{Atal:2019erb}
V.~Atal, J.~Cid, A.~Escriv{\`a} and J.~Garriga, \emph{{PBH in single field
  inflation: the effect of shape dispersion and non-Gaussianities}},
  \href{http://dx.doi.org/10.1088/1475-7516/2020/05/022}{\emph{JCAP} {\bf 05}
  (2020) 022}, [\href{http://arxiv.org/abs/1908.11357}{{\tt 1908.11357}}].

\bibitem{Escriva:2023qnq}
A.~Escriv{\`a} and C.-M. Yoo, \emph{{Primordial Black hole formation from
  overlapping cosmological fluctuations}},
  \href{http://dx.doi.org/10.1088/1475-7516/2024/04/048}{\emph{JCAP} {\bf 04}
  (2024) 048}, [\href{http://arxiv.org/abs/2310.16482}{{\tt 2310.16482}}].

\bibitem{Tomberg:2024chk}
E.~Tomberg, \emph{{Primordial black hole numbers: standard formulas and
  charts}},  \href{http://arxiv.org/abs/2408.09303}{{\tt 2408.09303}}.

\bibitem{Musco:2012au}
I.~Musco and J.~C. Miller, \emph{{Primordial black hole formation in the early
  universe: critical behaviour and self-similarity}},
  \href{http://dx.doi.org/10.1088/0264-9381/30/14/145009}{\emph{Class. Quant.
  Grav.} {\bf 30} (2013) 145009}, [\href{http://arxiv.org/abs/1201.2379}{{\tt
  1201.2379}}].

\bibitem{Blachier:2025iwk}
B.~Blachier and C.~Ringeval, \emph{{Friction in stochastic inflation}},
  \href{http://dx.doi.org/10.1088/1475-7516/2026/06/051}{\emph{JCAP} {\bf 06}
  (2026) 051}, [\href{http://arxiv.org/abs/2511.21388}{{\tt 2511.21388}}].

\bibitem{Joana:2025gqf}
C.~Joana and Z.-Y. Yuwen, \emph{{Primordial black holes from primordial
  voids}}, \href{http://dx.doi.org/10.1103/j3hw-d5cx}{\emph{Phys. Rev. D} {\bf
  113} (2026) 023518}, [\href{http://arxiv.org/abs/2510.11611}{{\tt
  2510.11611}}].

\bibitem{Pattison:2019hef}
C.~Pattison, V.~Vennin, H.~Assadullahi and D.~Wands, \emph{{Stochastic
  inflation beyond slow roll}},
  \href{http://dx.doi.org/10.1088/1475-7516/2019/07/031}{\emph{JCAP} {\bf 07}
  (2019) 031}, [\href{http://arxiv.org/abs/1905.06300}{{\tt 1905.06300}}].

\bibitem{Firouzjahi:2018vet}
H.~Firouzjahi, A.~Nassiri-Rad and M.~Noorbala, \emph{{Stochastic Ultra Slow
  Roll Inflation}},
  \href{http://dx.doi.org/10.1088/1475-7516/2019/01/040}{\emph{JCAP} {\bf 01}
  (2019) 040}, [\href{http://arxiv.org/abs/1811.02175}{{\tt 1811.02175}}].

\bibitem{Mishra:2023lhe}
S.~S. Mishra, E.~J. Copeland and A.~M. Green, \emph{{Primordial black holes and
  stochastic inflation beyond slow roll. Part I. Noise matrix elements}},
  \href{http://dx.doi.org/10.1088/1475-7516/2023/09/005}{\emph{JCAP} {\bf 09}
  (2023) 005}, [\href{http://arxiv.org/abs/2303.17375}{{\tt 2303.17375}}].

\bibitem{Artigas:2024ajh}
D.~Artigas, S.~Pi and T.~Tanaka, \emph{{Extended {\ensuremath{\delta}}N
  Formalism: Nonspatially Flat Separate-Universe Approach}},
  \href{http://dx.doi.org/10.1103/PhysRevLett.134.221001}{\emph{Phys. Rev.
  Lett.} {\bf 134} (2025) 221001}, [\href{http://arxiv.org/abs/2408.09964}{{\tt
  2408.09964}}].

\bibitem{Raveendran:2025pnz}
R.~N. Raveendran, \emph{{Validity of separate-universe approach in transient
  ultraslow-roll inflation}},
  \href{http://dx.doi.org/10.1103/2gd5-xv73}{\emph{Phys. Rev. D} {\bf 112}
  (2025) 103507}, [\href{http://arxiv.org/abs/2506.23571}{{\tt 2506.23571}}].

\bibitem{Briaud:2025ayt}
V.~Briaud, R.~Kawaguchi and V.~Vennin, \emph{{Stochastic inflation with
  gradient interactions}},
  \href{http://dx.doi.org/10.1088/1475-7516/2025/12/024}{\emph{JCAP} {\bf 12}
  (2025) 024}, [\href{http://arxiv.org/abs/2509.05124}{{\tt 2509.05124}}].

\end{thebibliography}\endgroup

\end{document}